\documentclass[sigconf, nonacm]{vldb/acmart}
\usepackage{caption}
\usepackage{subcaption}
\usepackage[linesnumbered,ruled,vlined]{algorithm2e}
\usepackage{setspace}
\usepackage{enumitem}
\usepackage{stfloats}
\usepackage{graphicx}
\usepackage{pifont}
\usepackage{url}
\usepackage{balance}
\usepackage[normalem]{ulem}
\usepackage[a-2b,mathxmp]{pdfx}
\usepackage{amsmath}
\usepackage{multicol}
\usepackage{multirow}
\usepackage{booktabs}
\usepackage{makecell}
\usepackage{utfsym}
\usepackage{xspace}
\usepackage{framed}

\usepackage{commons/code}
\usepackage{commons/colors}

\usepackage{amsthm}
\newtheorem{definition}{Definition}
\newtheorem{example}{Example}
\newtheorem{corollary}{Corollary}

\usepackage{balance}

\usepackage{color}
\usepackage{xcolor}
\usepackage{CJKutf8}

\newcommand{\blackding}[1]{\ding{\numexpr181+#1\relax}}

\newcommand\sysname{\ensuremath{\textsc{TxnSails}}\xspace}

\newif\ifextended\extendedfalse

\newcommand{\maintext}[1]{\ignorespaces\ifextended\relax\else#1\fi\ignorespaces}
\newcommand{\extended}[1]{\ifextended#1\else\relax\fi} 

\newcommand\vldbdoi{XX.XX/XXX.XX}

\newcommand\vldbavailabilityurl{https://github.com/dbiir/TxnSailsServer}

\begin{document}
\sloppypar
\extendedtrue
\maintext{\title{\sysname: Achieving Serializable Transaction Scheduling with Self-Adaptive Isolation Level Selection}}
\extended{\title{\sysname: Achieving Serializable Transaction Scheduling with Self-Adaptive Isolation Level Selection}}

\author{
 Qiyu Zhuang$^\dagger$, Wei Lu$^\dagger$, Shuang Liu$^\dagger$, Yuxing Chen$^\ddagger$, Xinyue Shi$^\dagger$  \\ Zhanhao Zhao$^\dagger$, Yipeng Sun$^\dagger$, Anqun Pan$^\ddagger$, Xiaoyong Du$^\dagger$\\
}
\affiliation{
\fontsize{10}{10}\textit{$^\dagger$ Renmin University of China}
    \qquad\fontsize{10}{10}\textit{$^\ddagger$ Tencent Inc.} 
 \\\fontsize{10}{10}{\it $^\dagger$\{qyzhuang, lu-wei, shuang.liu, xinyueshi, zhanhaozhao, yipengsun, duyong\}@ruc.edu.cn} 
 \\\fontsize{10}{10}{\it $^\ddagger$\{axingguchen, aaronpan\}@tencent.com}
}

\begin{abstract}
Achieving the serializable isolation level, regarded as the gold standard for transaction processing, 
is costly.
Recent studies reveal that adjusting specific query patterns within a workload can still achieve serializability even at lower isolation levels.
Nevertheless, these studies typically overlook the trade-off between the performance advantages of lower isolation levels and the overhead required to maintain serializability, potentially leading to suboptimal isolation level choices that fail to maximize performance.
In this paper, we present \sysname, a middle-tier solution designed to achieve serializable scheduling with self-adaptive isolation level selection. 
First, \sysname incorporates a unified concurrency control algorithm that achieves serializability at lower isolation levels 
with minimal additional overhead.
Second, \sysname employs a deep learning method to characterize the trade-off between the performance benefits and overhead associated with lower isolation levels, thus predicting the optimal isolation level. 
Finally, \sysname implements a cross-isolation validation mechanism to ensure serializability during real-time isolation level transitions. 
Extensive experiments demonstrate that \sysname outperforms state-of-the-art solutions by up to 26.7$\times$ and PostgreSQL's serializable isolation level by up to 4.8$\times$.
\end{abstract}
\maketitle




\pagenumbering{arabic}
\pagestyle{plain}

\section{Introduction}
Serializable isolation level (SER) is regarded as the gold standard for transaction processing due to its ability to prevent all forms of anomalies.
SER is essential in mission-critical applications, such as banking systems in finance and air traffic control systems in transportation, which require their data to be 100\% correct \cite{DBLP:journals/pvldb/ChenPLYHTLCZD24_TDSQL}.
However, it incurs expensive coordination overhead by configuring the RDBMS to SER \cite{DBLP:journals/pvldb/VandevoortK0N21}.
Despite significant efforts to alleviate this overhead~\cite {DBLP:conf/icde/LometFWW12,DBLP:conf/sosp/TuZKLM13,DBLP:conf/sigmod/YuPSD16}, maintaining a serial order of transactions to be scheduled remains a fundamental performance bottleneck.


In recent years, many studies have explored achieving SER by modifying applications while configuring the RDBMS to a low isolation level \cite{DBLP:journals/tods/KetsmanKNV22}. This approach is driven by two key reasons. First, some RDBMSs, such as Oracle 21c, cannot strictly guarantee SER and do not support the in-RDBMS modification of concurrency control, requiring application logic modifications to enforce it\extended{.
A more comprehensive list of RDBMSs and their supported isolation levels can be found in} \cite{DBLP:journals/pvldb/BailisDFGHS13}.
Second, RDBMSs typically offer better performance at lower isolation levels, such as read committed (RC) and snapshot isolation (SI), due to their more relaxed ordering requirements. Modifying applications to achieve SER while using a lower isolation level sometimes results in better performance compared to directly setting the RDBMS to SER~\cite{DBLP:conf/aiccsa/AlomariF15, DBLP:conf/icde/AlomariCFR08, DBLP:journals/pvldb/VandevoortK0N21}.

\begin{figure}[t]
    \centering
    \vspace{3mm}
    \begin{minipage}{0.95\linewidth}
        \centering
        \begin{subfigure}{\linewidth}
            \includegraphics[width=\linewidth]{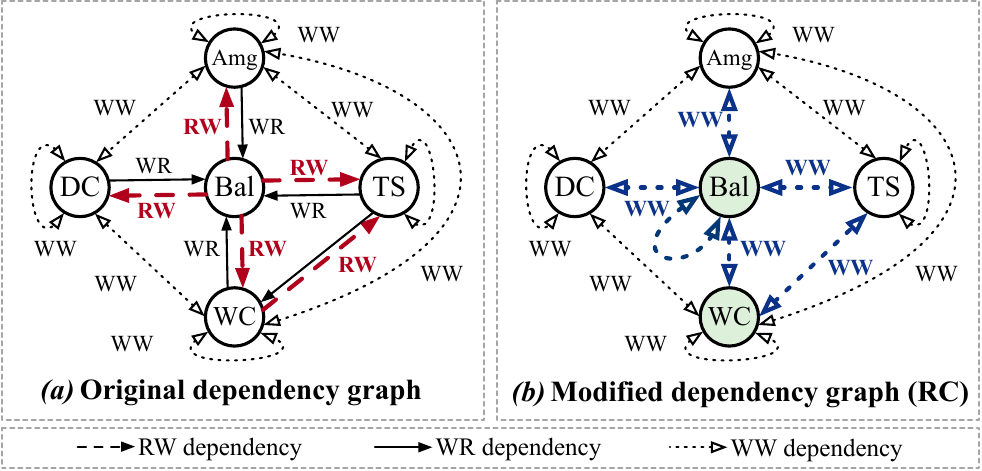}
        \end{subfigure}
    \end{minipage}
    \vspace{-2mm}
    \caption{Static dependency graphs of Smallbank}
    \label{fig:SmallBank}
    \vspace{-4mm}
\end{figure}

The main idea of existing works that can achieve SER under low isolation levels follows three steps: 
\blackding{1} Build a {\em static dependency graph} by analyzing the transaction templates, which are the abstraction of transaction logics in real-world applications~\cite{DBLP:journals/pvldb/VandevoortK0N21, DBLP:conf/icdt/VandevoortK0N22}.  
In this graph, each template is represented by a vertex, and the dependencies between templates, such as write-write (WW), write-read (WR), or read-write (RW), are depicted as edges.
\blackding{2} Configure the RDBMS to a low isolation level and identify {\em dangerous structures} that are permissible under the low isolation level but prohibited by SER according to the {\it static dependency graph}.
For example, under SI, a dangerous structure is characterized by two consecutive RW dependencies \cite{DBLP:conf/pods/Fekete05, DBLP:journals/sigmod/FeketeOO04}, while under RC, a single RW dependency constitutes the dangerous structure \cite{DBLP:conf/aiccsa/AlomariF15, DBLP:journals/pvldb/VandevoortK0N21}.
\blackding{3} Eliminate dangerous structures by modifying application logic, e.g., promoting reads to writes for certain SQL statements so that the RW dependencies are eliminated, and thus guarantees SER.

\begin{example}
\label{exa:oaofsb}
Consider the SmallBank benchmark \cite{DBLP:conf/icde/AlomariCFR08}, which consists of five transaction templates: Amalgamate (Amg), Balance (Bal), DepositChecking (DC), TransactSavings (TS), and WriteCheck (WC). Suppose the RDBMS is configured to RC.
As outlined in step \blackding{1}, the benchmark is modeled into a static dependency graph (shown in Figure \ref{fig:SmallBank}(a)). 
At step \blackding{2}, five dangerous structures (highlighted by red dashed arrows) are identified. These include the dependency from WC to TS and 4 dependencies from Bal to the other four templates. 
At step \blackding{3}, extra writes are introduced to convert RW dependencies to WW dependencies, thus eliminating the dangerous structures.
To achieve this, certain ``\textit{SELECT}'' statements are modified to ``\textit{SELECT ... FOR UPDATE}'' statements. 
\extended{Figure \ref{fig:sfu} illustrates the modification of Bal.}
\maintext{A more comprehensive illustration can be found in our supplementary material~\cite{TxnSails}.} 
For reference, the complete modified dependency graph, \extended{based on the original in Figure \ref{fig:SmallBank}(a), }is shown in Figure \ref{fig:SmallBank}(b).
\qed
\end{example}

\extended{
\begin{figure}[t]
    \centering
    \includegraphics[width=0.48\textwidth]{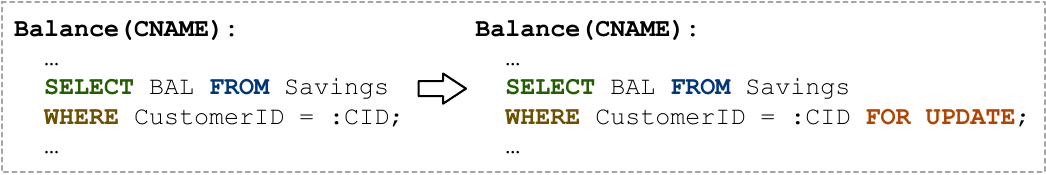}
    \vspace{-4mm}
    \caption{Modification of Bal transaction template}
    \label{fig:sfu}
    \vspace{-4mm}
\end{figure}
}

Thus far, existing studies~\cite{DBLP:journals/pvldb/BailisFFGHS14, DBLP:conf/pods/Fekete19, DBLP:conf/sigmod/BailisFHGS14, DBLP:journals/pvldb/FeketeGA09, DBLP:journals/tods/CahillRF09}  
have proposed promising solutions, enabling RDBMSs to achieve SER by operating at lower isolation levels while modifying specific query patterns within a workload. However, these studies exhibit two major shortcomings. 
First, the static modification of query patterns is inefficient.
These studies alter static SQL statements at the application level, converting certain read operations into write operations. This may result in unnecessary transaction conflicts. For instance, changing read operations to write operations may turn concurrent read-write operations into write-write conflicts in MVCC systems, thus significantly degrading transaction performance.
Second, these studies fail to address the key trade-off between the performance gains of lower isolation levels and the overhead needed to maintain SER, making it difficult to choose the optimal isolation level.
As shown in Figure \ref{fig:evaluation.dynamic} of \S\ref{sec:evaluation}, simply configuring the RDBMS to SER can sometimes outperform other methods. Additionally, as workloads evolve, the ideal isolation level may also shift, but existing studies lack the ability to adapt dynamically. 

In this paper, we present \sysname to address the aforementioned shortcomings with three key objectives:
\blackding{1} \sysname efficiently achieves SER under various low isolation levels.
\blackding{2} \sysname dynamically adjusts the optimal isolation level to maximize performance as the workload evolves.
\blackding{3} \sysname is designed to be general and adaptable across various RDBMSs, requiring no modifications to database kernels. To achieve this, \sysname is implemented as a middle-tier solution to enhance generalizability. 
However, implementing \sysname presents three major challenges. First, designing an approach that elevates various isolation levels to SER without introducing additional writes is a complex task. Second, determining the optimal isolation level requires accurately modeling the trade-offs between the performance benefits and serializability overhead associated with lower isolation levels, which is particularly challenging in dynamic workloads. Third, as workloads evolve, the optimal isolation level may need to adapt over time, making it essential to design an efficient and reliable mechanism for transitioning between isolation levels.
To address these challenges, we propose the following key techniques.

\textbf{(1) Efficient middle-tier concurrency control algorithm ensuring SER for each low isolation level (\S\ref{design-1}).} 
We propose a runtime, fine-grained approach that operates on individual transactions rather than transaction templates, ensuring that the execution of transactions meets the requirements of SER. This approach is inspired by the theorem that a schedule is serializable if it does not contain two transactions, $T_i$ and $T_j$, where $T_j$ commits before $T_i$, but there is a dependency from $T_i$ to $T_j$ \cite{DBLP:conf/aiccsa/AlomariF15}.
Building on this theorem, we introduce a unified concurrency control algorithm to ensure SER. 
The algorithm tracks transactions with their templates involved in specific RW dependency within a static dependency graph. 
It detects the runtime dependencies and schedules the commit order to align with their dependency order. If necessary, it will abort a transaction to guarantee SER.

\textbf{(2) Self-adaptive isolation level selection mechanism (\S\ref{design-2}).} 
%
Rather than directly quantifying a cost model to balance the trade-off between the performance benefits of lower isolation levels and the overhead required to maintain serializability, we employ a learned model leveraging graph neural networks~\cite{DBLP:journals/corr/BrunaZSL13} and message-passing techniques~\cite{DBLP:conf/icml/GilmerSRVD17} to predict the optimal isolation level for a given workload.
Our observations reveal that the performance of various isolation levels and the overhead of achieving SER are closely influenced by two critical factors: the data access dependencies between transactions and the data access distribution within transactions. To capture these relationships, we model workload features as a graph, where vertices represent individual transaction features and edges denote data access dependencies between transactions.
Building on this insight, we propose a graph-based model for dynamic workloads that predicts the optimal isolation level using real-time workload features.
To the best of our knowledge, \textit{\sysname is the first work to enable self-adaptive isolation level selection for dynamic workloads.}

\textbf{(3) Cross-isolation validation mechanism that enables efficient transitions and serializable scheduling (\S\ref{design-3}).} 
The optimal isolation level should adapt as the workload evolves. When the RDBMS decides to change the isolation level, new transactions must be executed under this updated isolation level. Although existing approaches can achieve SER when all transactions use a unified low isolation level, they fail to ensure SER when transactions operate under different isolation levels. This is because varying isolation levels can introduce new dangerous structures that may violate the requirements of SER.
To address this issue, we identify dangerous structures across different isolation levels and propose a cross-isolation validation mechanism that can prevent the occurrence of these structures during transitions without causing significant system downtime.
We prove the correctness of the cross-isolation validation mechanism in \S\ref{sec:proof.switch}. 


We have conducted extensive evaluations on SmallBank \cite{DBLP:conf/icde/AlomariCFR08}, TPC-C \cite{TPCC}, and YCSB+T \cite{DBLP:conf/cloud/CooperSTRS10} benchmarks. The results show that \sysname can adaptively select the optimal isolation level for dynamic workloads, achieving up to a 26.7$\times$ performance improvement over other state-of-the-art methods and up to a 4.8$\times$ performance boost compared to SER provided by PostgreSQL.

\section{preliminaries}
RDBMSs typically offer several isolation levels; in this paper, we focus on the three most commonly used: serializable (SER), snapshot isolation (SI), and read committed (RC).
In this section, we first discuss transaction templates.
We then present the dangerous structures under SI and RC, respectively.
We finally define the vulnerable dependencies that build the foundation of our approach.

\subsection{Transaction Templates}



A transaction template is an abstraction of application logic that consists of predefined SQL statements with parameter placeholders. 
Take the Amalgamate template in Example \ref{exa:oaofsb} as an example, 
which facilitates the transfer of funds from one customer to another. It first reads the balances of the checking and savings accounts of customer $N_1$, then sets them to zero. Finally, it increases the checking balance for $N_2$ by the sum of $ N_1$'s previous balances. In this context, $N_1$ and $N_2$ serve as parameter placeholders. This modular structure ensures readability and flexibility, allowing the transaction template to be reused across various contexts.
When a customer initiates a transaction at runtime, the application fills the placeholders with actual data. The complete transaction is then executed in the RDBMS, finalizing the business logic.

For better clarity, we use $\mathcal{T}_i$ to denote a transaction template and $T_i$ to denote a transaction generated by $\mathcal{T}_i$.

\subsection{Dangerous Structures\label{sec:back.vd}}
The dependencies between two concurrent transactions, $T_i$ and $T_j$, operating on the same item $x$, are classified as follows.

\begin{itemize}[leftmargin=*]
    \item $T_i \xrightarrow{ww} T_j$: $T_i$ writes a version of data item $x$, and $T_j$ writes a later version of $x$.
    \item $T_i \xrightarrow{wr} T_j$: $T_i$ writes a version of data item $x$, and $T_j$ reads either the version written by $T_i$ or a later version of $x$.
    \item $T_i \xrightarrow{rw} T_j$: $T_i$ reads a version of data item $x$, and $T_j$ writes a later version of $x$.
\end{itemize}

\begin{definition}[SI dangerous structure~\cite{DBLP:journals/pvldb/PortsG12}]
    \label{def:si}
    Under SI, two consecutive RW dependencies: $T_i \xrightarrow{rw} T_j \xrightarrow{rw} T_k$ are considered as an SI dangerous structure, where $T_i$ and $T_j, T_j$ and $T_k$ are concurrent transactions, respectively. 
    \qed
\end{definition}

\begin{definition}[RC dangerous structure~\cite{DBLP:journals/pvldb/GanRRB020, DBLP:conf/aiccsa/AlomariF15}]
    \label{def:rc}
    Under RC, an RW dependency: $T_i \xrightarrow{rw} T_j$ is considered as an RC dangerous structure, where $T_i$ and $T_j$ are concurrent transactions. 
    \qed
\end{definition}


When it comes to transaction templates, the dependencies between two transaction templates, $\mathcal{T}_i$ and $\mathcal{T}_j$, are defined as follows: (1) $\mathcal{T}_i \xrightarrow{ww} \mathcal{T}_j$ if $\mathcal{T}_i$ and $\mathcal{T}_j$ write the same data set (e.g., relation) in sequence; (2) $\mathcal{T}_i \xrightarrow{wr} \mathcal{T}_j$ if $\mathcal{T}_i$ writes and $\mathcal{T}_j$ reads the same data set in sequence; (3) $\mathcal{T}_i \xrightarrow{rw} \mathcal{T}_j$ if $\mathcal{T}_i$ reads and $\mathcal{T}_j$ writes the same data set in sequence.


\begin{definition}[Static SI dangerous structure~\cite{DBLP:conf/sigmod/CahillRF08, alomari2008serializable}]
    \label{def:sta_si}
    In a static dependency graph, two consecutive edges $\mathcal{T}_i \xrightarrow{rw} \mathcal{T}_j$, $\mathcal{T}_j \xrightarrow{rw} \mathcal{T}_k$ are deemed to constitute a static SI dangerous structure.
    \qed
\end{definition}

\begin{definition}[Static RC dangerous structure~\cite{DBLP:conf/aiccsa/AlomariF15, DBLP:conf/icdt/VandevoortK0N22}]
    \label{def:sta_rc}
    In a static dependency graph, an edge $\mathcal{T}_i \xrightarrow{rw} \mathcal{T}_j$ is deemed to constitute a static RC dangerous structure.
    \qed
\end{definition}

\begin{theorem}[\cite{alomari2009ensuring}]
\label{the:RC}
If a static dependency graph contains no SI (resp. RC) static dangerous structures, then scheduling the transactions generated by the corresponding transaction templates achieves SER when the RDBMS is configured to SI (resp. RC).
\qed
\end{theorem}


Theorem \ref{the:RC} serves as the foundation for existing approaches to achieving SER while the RDBMS is configured to SI/RC. However, these approaches are static and coarse-grained, leading to the incorrect identification of many non-cyclic schedules. This, in turn, causes a significant number of unnecessary transaction rollbacks.


\subsection{Vulnerable Dependency}
\begin{definition}[Static vulnerable dependency]
    \label{def:static_vul}
    The static vulnerable dependency is defined as $\mathcal{T}_j \xrightarrow{rw} \mathcal{T}_k$ in chain $\mathcal{T}_i \xrightarrow{rw} \boxed{\mathcal{T}_j \xrightarrow{rw} \mathcal{T}_k}$ under SI, and $\boxed{\mathcal{T}_i \xrightarrow{rw} \mathcal{T}_j}$ under RC, respectively.  
    \qed
\end{definition}

\begin{definition}[Vulnerable dependency]
    \label{def:vul}
    The vulnerable dependency is defined as $T_j \xrightarrow{rw} T_k$ in chain $T_i \xrightarrow{rw} \boxed{T_j \xrightarrow{rw} T_k}$ under SI, and $\boxed{T_i \xrightarrow{rw} T_j}$ under RC, respectively.
    \qed
\end{definition}

\begin{theorem}[\cite{DBLP:conf/aiccsa/AlomariF15}] 
\label{the:vulnerable}
For any vulnerable dependency $T_i\xrightarrow{rw} T_j$, if $T_i$ commits before $T_j$, then the scheduling achieves SER.
\qed
\end{theorem}

Theorem \ref{the:vulnerable} forms the basis of our dynamic, fine-grained approach to achieving serializable scheduling. 
Compared to existing approaches, our approach neither introduces unnecessary writes nor misjudges cyclic schedules, thus preventing unwarranted transaction rollbacks.

\section{Overview of \sysname}
\label{sec:overview}

\begin{figure}[t]
    \centering
    \includegraphics[width=0.47\textwidth]{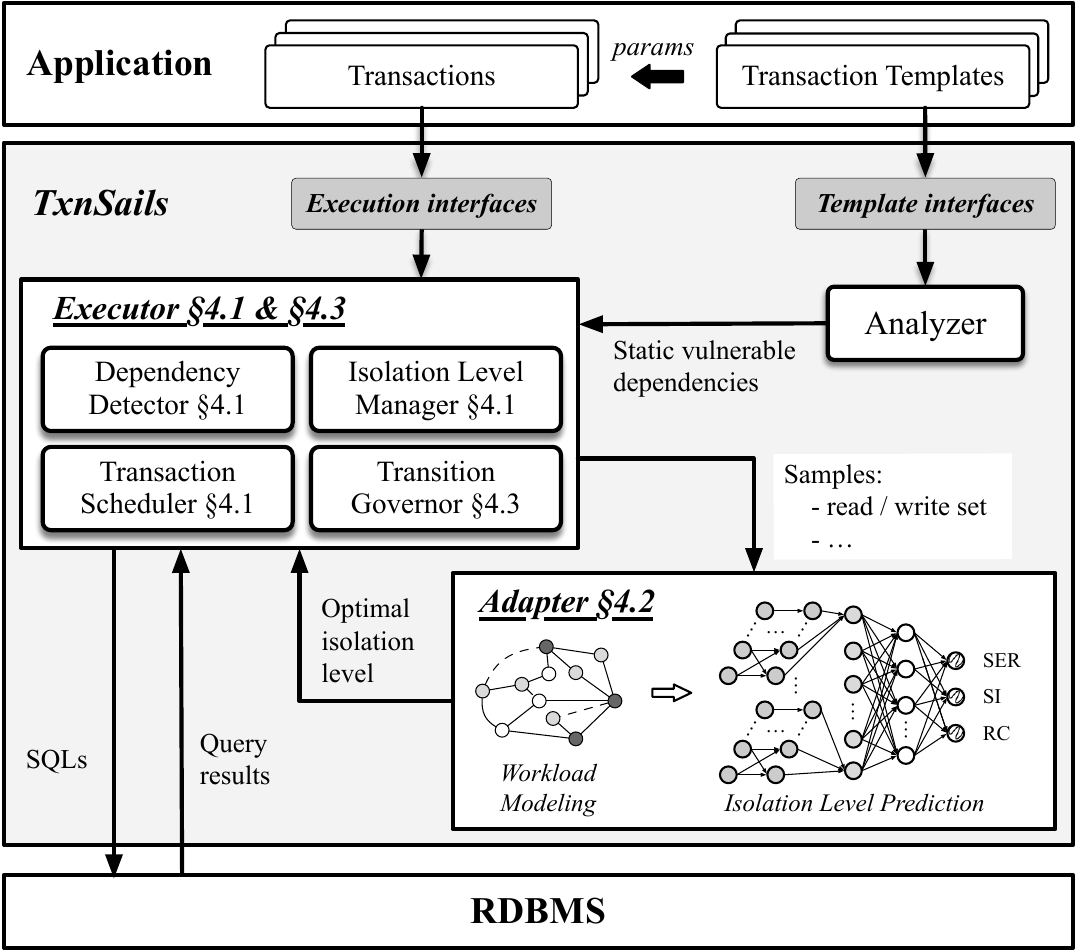}
    \vspace{-2mm}
    \caption{An overview of \sysname}
    \Description{A concise description of the figure for screen readers.}
    \label{fig:overview}
    \vspace{-4mm}
\end{figure}

\sysname works in the middle tier between the application tier and the database tier, designed to \blackding{1} ensure SER when transactions operate under a low isolation level without introducing additional writes; \blackding{2} select the optimal isolation level for dynamic workloads adaptively; \blackding{3} constantly keep SER during the isolation level transition. 
An overview of \sysname is depicted in Figure~\ref{fig:overview}. It comprises three main components: \textit{Analyzer}, \textit{Executor}, and \textit{Adapter}.

\noindent\textbf{Analyzer.} 
{
\textit{Analyzer} provides \textit{template interfaces} for template registration and analysis. 
Before \sysname executes any transaction from the application, \textit{Analyzer} builds the static dependency graph for the transaction templates and identifies all the static vulnerable dependencies for each low isolation level according to Definition \ref{def:static_vul}. It then sends the static vulnerable dependencies to \textit{Executor}.




\noindent\textbf{Executor.} 
\textit{Executor} provides \textit{execution interfaces} for applications to execute transactions. It ensures SER when transactions operate either at a single low isolation level or during the isolation level transition. 
There are four core modules: \textit{Isolation Level Manager (ILM)}, \textit{Dependency Detector (DD)}, \textit{Transaction Scheduler (TS)}, and \textit{Transition Governor (TG)}. 
(1) ILM stores the static vulnerable dependencies, and before any transaction $T$ starts, it identifies whether $T$ involves any static vulnerable dependencies. If the template of $T$ does not involve static vulnerable dependencies, \textit{Executor} sends $T$ directly to the RDBMS for execution; otherwise, ILM triggers DD that identifies vulnerable dependencies of $T$.
(2) DD monitors the reads and writes of $T$, detecting any runtime vulnerable dependencies between $T$ and other transactions. If $T$ is involved in any vulnerable dependencies, TS is triggered. 
(3) TS attempts to ensure that the commit and vulnerable dependency orders remain consistent between $T$ and other transactions. If the consistency cannot be guaranteed, $T$ is blocked or aborted; otherwise, $T$ proceeds to commit.
(4) TG ensures SER during the transition between two isolation levels. It follows a new corollary, which extends Theorem \ref{the:vulnerable} 
to any two transactions, $T_i$ and $T_j$, executing under different isolation levels. 
The proof of correctness during the isolation level transition is detailed in \S\ref{sec:proof.switch}.

\noindent\textbf{Adapter.} \textit{Adapter} models the trade-off between performance benefits of low isolation levels and additional serializability overhead outside the database kernel. It predicts the optimal isolation level when the workloads evolve. Initially, a dedicated thread is introduced to continuously sample aborted/committed transactions using Monte Carlo sampling \cite{DBLP:journals/entropy/ZhouJLWLG24}, capturing the read/write data items. 
After collecting a batch of transaction samples, \textit{Adapter} predicts the optimal isolation level for future workloads based on the characteristics of the batch. The prediction process consists of two steps: \textit{Workload Modeling (WM)} and \textit{Isolation Level Prediction (ILP)}. 
WM extracts performance-related features and models the workload as a graph.
In this graph, each vertex represents a runtime transaction, with its features capturing the transaction context, such as the number of data items in the read and write sets.
Each edge represents an RW or WW operation dependency between transactions.
Following WM, ILP embeds the workload graph into a high-dimension vector using graph neural network \cite{DBLP:journals/corr/BrunaZSL13} and message passing techniques \cite{DBLP:conf/icml/GilmerSRVD17}, and then translates the vector into three possible labels: RC, SI, or SER. 
The label with the highest value, as determined by our model, indicates the most efficient isolation level.

For reference, the detailed implementation of the interfaces and the three core components are provided in \S\ref{implementation}.

\section{Design of \sysname}
In this section, we provide the detailed design of \sysname. We first introduce the middle-tier concurrency control mechanism that achieves serializable scheduling when the RDBMS is configured to a low isolation level (\S\ref{sec:validation-based-cc}). Then, we present a self-adaptive isolation level selection approach, which can predict the optimal future isolation level (\S\ref{design-2}). Lastly, we introduce the cross-isolation validation mechanism that ensures serializable scheduling during the isolation level transition (\S\ref{design-3}). 

\subsection{Middle-tier Concurrency Control \label{design-1}}
\label{sec:validation-based-cc}
Existing approaches ensure serializability at low isolation levels by statically introducing additional write operations. However, these approaches reduce concurrency and increase overhead.
To overcome these limitations, \sysname introduces a middle-tier concurrency control algorithm, which dynamically validates runtime dependencies and schedules their commit order. 
In particular, \sysname focuses exclusively on vulnerable dependencies identified by the \textit{Analyzer} and employs a lightweight validation mechanism to mitigate overhead further.
\subsubsection{Transaction lifecycle} 

The lifecycle of transactions in the middle tier is divided into three phases: execution, validation, and commit phases.
(1) In the execution phase, \sysname establishes a database connection with a specific isolation level, which is not adjusted until the transaction is committed or aborted. 
Following the RDBMS transaction execution, \sysname stores the read/write data items in the thread-local buffer that may induce the vulnerable dependencies;
(2) In the validation phase, \sysname acquires validation locks for data items stored in the buffer. Then, it detects the dependencies among them and aims to schedule the commit order consistent with the identified dependency order.
A more detailed description of the validation phase will be given in \S\ref{sec:design:cc:validation}; 
(3) In the commit phase, \sysname applies modifications to the database and subsequently releases the validation locks.

\subsubsection{Validation phase\label{sec:design:cc:validation}} \sysname performs two key tasks in the validation phase: (1) detecting vulnerable dependencies; (2) scheduling the commit order consistent with the dependency order. 
To achieve this, we explicitly add a \textit{version} column to the schema, which is incremented after every update.
We trace the dependency orders by comparing the versions of data items. \maintext{
Due to space constraints, we provide a concise overview of the validation algorithm, while leaving the detailed pseudocode in our technique report~\cite{TxnSails}.
}
\extended{
Algorithm~\ref{alg.transaction} shows the detailed algorithm. 
}



\extended{
\begin{algorithm}[t]
    \caption{Middle-tier concurrency control algorithm}
    \small
    \label{alg.transaction}
    \SetKwInOut{KwIn}{Input}
    \SetKwFunction{Validate}{Validate}
    \SetKwProg{Fn}{Function}{:}{}

    \Fn{\Validate{T, conn}} {
        \KwIn{T, transaction requiring validation; \\ conn, a connection under RC}
        \tcp{Acquire the validation locks on data items}
        \For {r \textbf{in} T.vread\_set $\cup$ T.vwrite\_set} {
            res := TryValidationLock(r.key, T.tid, r.type) \\
            \While {res \textbf{is} WAIT} {
                res := TryValidationLock(r.key, T.tid, r.type) \\
            }
            \If {res \textbf{is} ERROR} {
                \Return {ERROR} 
            }
        }
        \tcp{Check the version of data items in the read set}
        \For{r \textbf{in} T.vread\_set} {
            version := 0\\
            entry := VLT.get\_lock\_entry(r.key) \\
            \If {entry.version > 0} { 
                \tcp{get the latest version from version cache}
                version := entry.version \\
            }
            \Else { 
            \tcp{fetch the latest version from DBMS}
                version := conn.get\_version(r.key) \\
                entry.version := version\\
            }
            \If {version \textbf{is not} r.version} {
                \Return {ERROR} \\
            }
        }
        \Return{SUCCESS}\\
    }

    \SetKwFunction{Commit}{Commit}
    \SetKwProg{Fn}{Function}{:}{}

    \Fn{\Commit{T, sess}} {
        \KwIn{sess, session for transaction execution;}
        sess.commit(T)\\
        \For{r \textbf{in} T.vwrite\_set} {
            entry := VLT.get\_lock\_entry(r.key) \\
            entry.version = r.version \\
        }
        \For {r \textbf{in} T.vread\_set $\cup$ T.vwrite\_set} {
            ReleaseValidateLock(r.key) \\
        }
    }
\end{algorithm}
}

For both RC and SI levels, we detect vulnerable dependencies based on those defined in Definition~\ref{def:vul}.  
During validation, the transaction first requests \textit{Shared} locks for items in the read set and \textit{Exclusive} locks for items in the write set\extended{ (lines 2-7)}. Specifically, before transaction $T_i$ commits,  the validation phase is performed in two key steps:

(1) $T_i$ checks each data item in its write set to determine if any concurrent transaction $T_j$ is reading the same data item and undergoing validation. 
We achieve this via the validation locks. 
If any lock request fails, indicating $T_j$ exists, an RW dependency is detected. 
In such a case, the failed lock request should be appended in the corresponding lock's waitlist, making $T_i$ wait until $T_j$ commits, ensuring consistency between dependency and commit orders.  
If no concurrent transactions in the validation phase are reading the same item, $T_i$ proceeds to commit and create a new data version. 

(2) $T_i$ checks each data item in its read sets and compares the version of each read item in the thread-local buffer with their latest version maintained\extended{ (lines 10-15)}. If a newer version is found, indicating an RW dependency from the current transaction $T_i$ to a committed transaction, say $T_j$, then $T_i$ is aborted to 
ensure the consistency of commit and dependency orders\extended{ (lines 16-17)}. Moreover, comparing data item versions in the local buffer with the latest versions in the RDBMS can introduce additional interactions between the database middleware and the underlying database, imposing overhead on system and network resources. To alleviate this burden, \sysname employs a caching mechanism in the middle-tier memory to store the latest versions of data items. By enabling rapid retrieval of the most recent versions, this approach significantly reduces validation overhead and enhances overall efficiency. 

In the above steps, we ensure the commit order in the middle tier is consistent with the dependency order. Subsequently, we schedule the actual commit order in the RDBMS consistent with the commit order in the middle tier. Middle-tier consistency is achieved by aborting or blocking transactions when detecting a vulnerable dependency. We ensure that the RDBMS layer consistency is achieved by releasing validation locks only after the transaction has been successfully committed in the database\extended{ (lines 20-25)}. Based on this, if two concurrent transactions access the same data item (with one writing and the other reading or writing), they cannot both enter the validation phase simultaneously. One transaction must complete validation and commit before the other can proceed, ensuring a correct and consistent commit order in the RDBMS.

To enable efficient and accurate validation, \sysname leverages a validation lock table (VLT) to maintain metadata for each data item. Each data item is assigned a hash value computed using the collision-resistant hash function $\mathcal{H}$, and a corresponding entry is stored in VLT. 
Data items with the same hash value are stored in the same bucket and organized as a linked list. When an entry with key \( x \) is accessed, \sysname first determines the appropriate bucket using \( \mathcal{H}(x) \) and then traverses it to locate the specific entry. 
Each hash entry $e$ comprises five fields: (1) $e.Type$, the type of locks acquired, which can be \textit{None}, \textit{Shared (SH)}, and \textit{Exclusive (EX)}; (2) $e.LockNum$, the number of currently held locks; (3) $e.WaitList$, a list of transactions waiting to acquire locks; (4) $e.LatestVersion$, the most recent committed version of the data item; and (5) $e.Lease$, the timestamp indicating the garbage collection time. 

\begin{figure}[t]
    \centering
    \includegraphics[width=0.46\textwidth]{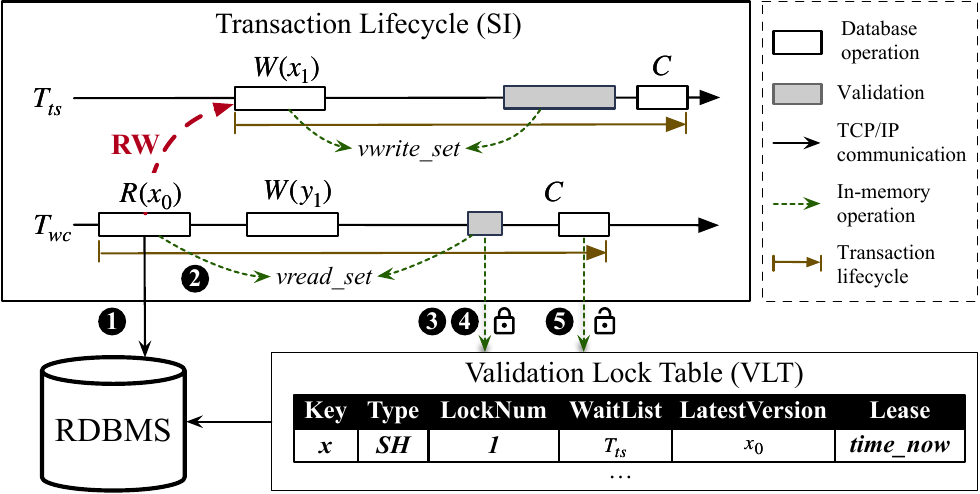}
    \vspace{-3mm}
    \caption{Transaction processing in \sysname}
    \label{fig:transaction_lifecycle}
    \vspace{-4mm}
\end{figure}

\begin{example}
    Take Figure~\ref{fig:transaction_lifecycle}, which provides a concise depiction of transaction processing, as an example. Recall that there exists a static vulnerable dependency $\mathcal{T}_{wc} \xrightarrow{rw} \mathcal{T}_{ts}$ in Smallbank when the RDBMS is set to SI (Figure \ref{fig:SmallBank}). Thus, it is necessary to detect the read operation of $T_{wc}$ and the write operation of $T_{ts}$. 
    In the execution phase, after the RDBMS execution (\blackding{1}), $T_{wc}$ stores the data item \textit{x} in its \textit{vread\_set} and $T_{ts}$ stores \textit{x} in its \textit{vwrite\_set} (\blackding{2}). 
    In the validation phase of $T_{wc}$, it acquires the shared validation lock on \textit{x} (\blackding{3}) and retrieves the latest version of \textit{x} from either VLT or the RDBMS (\blackding{4}). 
    While in the validation phase of $T_{ts}$, it requests the exclusive validation lock on \textit{x} and is blocked until $T_{wc}$ releases the lock. 
    Finally, in the commit phase, $T_{wc}$ releases the validation lock on \textit{x} (\blackding{5}). This ensures that the commit order of the two transactions is consistent with the dependency order, thereby guaranteeing SER when they operate under SI.
    \qed
\end{example}

\subsubsection{Discussion} 
To optimize memory usage in VLT, \sysname incorporates an efficient garbage collection algorithm to evict cold entries. Upon accessing an entry \( e \) in the VLT, \sysname updates \( e.Lease \) with a future timestamp, then traverses the corresponding bucket to identify and remove outdated, unused entries where \( Lease \) falls behind than the current timestamp and \( Type \) is \textit{None}. 
Additionally, to prevent entries in infrequently accessed buckets from persisting indefinitely, a background thread periodically scans these long-unused buckets and evicts outdated entries, ensuring efficient memory management across the system. 

We note that range queries with predicates may potentially introduce phantom reads anomaly. 
For phantom reads, the definition of vulnerable dependency remains applicable, which enables \sysname to detect and prevent this anomaly. 
The only difference is that we need to implement a larger granule validation lock, such as interval or table locks, to enable detecting the dependencies between predicates. 
As various coarse-grained locking techniques, such as SIREAD locks in PostgreSQL and gap locks~\cite{DBLP:conf/vldb/Lomet93}, already exist, we opt to implement the coarse-grained validation locks using these methods and exclude locking optimization from our paper to focus on efficient isolation level adaptation.

\subsection{Self-adaptive Isolation Level Selection \label{design-2}}\label{sec:self-adaptive_section}



Selecting optimal isolation levels for all transactions in a workload while maintaining SER is challenging, as we need to balance the overhead and performance gain in different isolation levels.  
Inspired by existing approaches
that effectively conduct workload prediction using neural networks~\cite{DBLP:conf/icde/ZhengZLYCPD24, DBLP:conf/sigmod/MaAHMPG18, DBLP:journals/pvldb/YuZSY24}, 
we propose a neural-network-based isolation level prediction approach, which predicts the future optimal isolation level based on the current workload features. 
The main challenges are effective workload feature selection and representation. 
Towards this end, \sysname adopts transaction dependency graphs to capture the workload features and adopts a graph classification model to perform self-adaptive isolation level prediction. 

\subsubsection{Graph construction.} 

To extract the complex features of concurrent transactions, \sysname proposes a graph-structured workload model, which is composed of three matrixes: a vertex matrix $V$, an edge index matrix $E$, and an edge attribute matrix $A$. Formally, a workload graph is defined as $G=(V, E, A)$, where each row in $A$ represents the feature vector of an operator, each entry $e_{ij}$ in $E$ signifies the relationship between $v_i$ and $v_j$, and each row in $A$ represents the feature vector of an edge. 

\sysname dynamically constructs the runtime workload graph by sampling transactions adhering to Monte Carlo sampling. Each transaction in the batch is mapped to a vertex $v_i$, and its feature vector $V_i$ is generated by extracting the number of data items in its read set and write set.
For each vertex pair $(v_i, v_j)$, if a data dependency exists between them, i.e., their read and write sets intersect, \sysname adds an edge entry $e_{ij}$ into the edge index matrix $E$. For each edge $e_{ij}$, \sysname extracts the data dependency type and the involved relations to generate its attribute $A_{e_{ij}}$. The data dependency type for $e_{ij}$ can be either RR, RW/WR, or WW. For example, if two transactions' write sets intersect, there is a WW dependency between them. Given that the number of relations and dependency types are fixed, one-hot encoding is employed to represent these two features within the attribute matrix.

\subsubsection{Graph embedding and isolation prediction} 

Predicting the optimal isolation strategy for the future workload using the constructed graph-structure model $G=(V, E, A)$ is challenging due to its complex structures and dynamic and high-dimensional features, which require capturing both local and global dependencies. Heuristic methods rely on manually crafted rules that lack generalizability, while traditional machine learning models are deficient in leveraging relational information encoded in the vertexes and edges, losing critical structural context. To address these challenges, we use a graph classification model that learns graph-level representations by aggregating node features through multiple layers of neural network-based convolutions.

\begin{figure}[t]
    \centering
    \begin{minipage}{0.99\linewidth}
        \centering
        \begin{subfigure}{1.0\linewidth}
            \includegraphics[width=\linewidth]{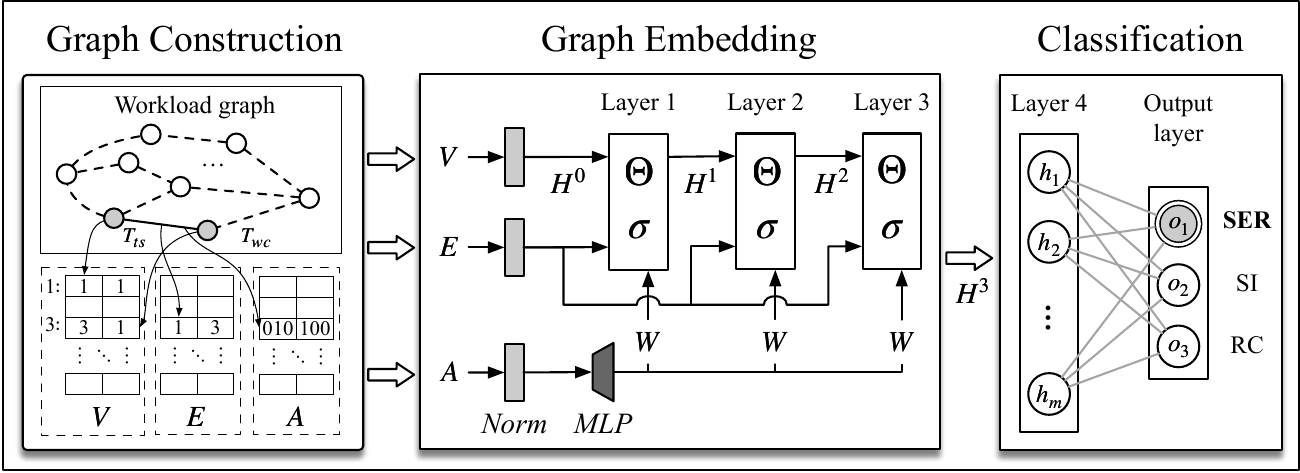}
        \end{subfigure}
    \end{minipage}
    \vspace{-2mm}
    \caption{Graph-based isolation level selection model}
    \label{fig:graph_learning}
    \vspace{-6mm}
\end{figure}

As shown in Figure~\ref{fig:graph_learning}, our graph model comprises two parts. First, we use a \textit{Graph Embedding Network} to learn and aggregate both vertex and edge features, producing node-level embedded matrix $H$ that encodes the local structure and attribute information of the graph. Second, to predict the optimal isolation strategy for the workload, we use a \textit{Graph Classification Network} that learns the mapping from the embedded matrix $H$ to perform the end-to-end graph classification to predict the optimal isolation strategy.

The \textit{Graph Embedding Network} is constructed with a three-layer architecture, where each layer applies a convolution operation to update node representations. This process integrates node and edge features through a dynamic aggregation mechanism~\cite{DBLP:journals/pvldb/ZhouSLF20,DBLP:conf/pkdd/FurutaniSAHA19}. At each layer, an edge network maps the input edge features into higher-dimensional convolution kernels via a multi-layer perception (MLP), as shown in Eq.(\ref{eq: gnn}). This mapping dynamically transforms edge attributes into weights, which are then used during the node aggregation step. The convolution operation produces updated node embeddings for each node $v_i$ using Eq.~\ref{eq: gnn}, where $\mathcal{N}(v_i)$ represents the neighbors of node $v_i$, $W^{(l)}_{e_{ij}}$ is the edge-specific weight, and $\sigma$ denotes the active function (i.e., ReLU). Through this layer-wise propagation, the embedding module produces $H$, a set of node-level embeddings that encode the graph information.

\begin{equation}
    \small
    \begin{cases}
    W^{(l)} = f^{(l)}(A) = MLP(A) \\
    H^{(l)}_{v_i} = \sigma \left( \underset{v_j \in \mathcal{N}(v_i)}{\max} \left( W^{(l)}_{e_{ij}} \cdot H_{v_j}^{(l-1)} \right) \right)
    \end{cases}
    \label{eq: gnn}
\end{equation} 

The \textit{Graph Classification Network} takes the node embeddings $H$ as inputs and passes them through two fully connected layers. The first layer applies a ReLU activation function to enhance nonlinearity. The second layer implements a softmax activation function and outputs a three-dimensional vector, with each field representing the probability of the isolation level being optimal.

\subsubsection{Data collection and labeling. } Our modeling approach is somewhat general and not designed specifically for specific workloads. However, in practice, we train the model separately for each type of benchmark for efficiency considerations. Taking YCSB+T as an example, we generate lots of random workloads with varying read/write ratios and key distributions. 
Each workload is executed under each isolation level for 10 seconds, with sampling intervals of 1 second, and the optimal isolation level is labeled based on throughput. 
We follow the same process for data collection and labeling in Smallbank and TPC-C. 


\subsubsection{Model training.} In \sysname, we train the embedding and prediction network together and use cross-entropy loss for multi-class classification. 
Backpropagation involves calculating the gradients of the loss function concerning the parameters of the graph model. First, the gradient is computed for the output layer. Then, using the chain rule, these gradients are propagated backward through the whole network, updating the parameters of each layer. For embedding layers, this process includes computing gradients for both vertex features and transformation matrices derived from edge attributes.
The model training overhead after the transaction template change is insignificant because our model parameters are 550k. The model can be retrained asynchronously, and during the retraining period, it does not affect transaction execution.


\subsection{Cross-isolation Validation \label{design-3}} \label{sec:switch_mechanism}
If the predicted optimal isolation level changes, \sysname will adapt from the previous isolation level $I_{old}$ to the optimal isolation level $I_{new}$. We design a cross-isolation validation mechanism to guarantee SER during the isolation level transition.

\begin{figure}[t]
    \centering
    \includegraphics[width=0.47\textwidth]{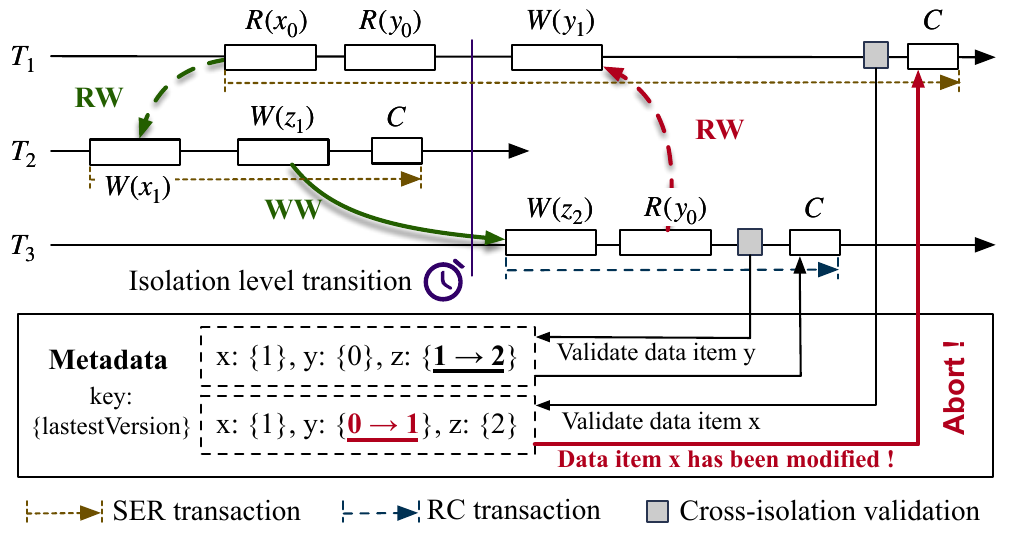}
    \vspace{-4mm}
    \caption{Cross-isolation validation}
    \label{fig:cross-isolation}
    \vspace{-4mm}
\end{figure}

\begin{example}
    \label{exa:cross-isolation}
    Figure~\ref{fig:cross-isolation} illustrates non-serializable scheduling during the transition from SER to RC after $T_2$ commits, making $T_1$ and $T_2$ operate under SER while $T_3$ operates under RC. In this scenario, $T_1$ is expected to be aborted to ensure SER. However, existing RDBMSs do not handle dependencies between transactions under different isolation levels, allowing $T_1$ to commit successfully, leading to non-serializable scheduling.
    Note that when transactions $T_1$, $T_2$, and $T_3$ are all executed under SER, the concurrency control in RDBMS prevents such non-serializable scheduling. 
    \qed
\end{example}

We need to explicitly consider the situations of cross-isolation transitions to ensure the correct transaction execution during the process. 
A straightforward approach is to wait for all transactions to complete under the previous isolation level before making the transition. In the example above, this would mean blocking $T_3$ until $T_1$ commits. 
However, it can result in prolonged system downtime, especially when there are long-running uncommitted transactions. Another possible approach is to abort these uncommitted transactions and retry them after the transition, which leads to a high abort rate. 
To mitigate these negative impacts, \sysname employs a cross-isolation validation (CIV) mechanism that ensures serializability and allows for non-blocking transaction execution without a significant increase in aborts. 
Specifically, we extend the vulnerable dependency under the single isolation level in Definition~\ref{def:vul} to the cross-isolation vulnerable dependency, defined as follows:
\begin{definition}[Cross-isolation vulnerable dependency]
    \label{def:transition_vul}
    The cross-isolation vulnerable dependency is defined as $T_j \xrightarrow{rw} T_k $ in chain $T_i \xrightarrow{rw} T_j \xrightarrow{rw} T_k$ where three transactions can execute under two different isolation levels.
    \qed
\end{definition}

We extend Theorem \ref{the:vulnerable} to obtain corollary~\ref{cor:transition_ser} and prove it in \S\ref{sec:proof.switch}. 
\begin{corollary}
    \label{cor:transition_ser}
    For any cross-isolation vulnerable dependency $T_i \xrightarrow{rw} T_j$, if $T_i$ commits before $T_j$, then the transaction scheduling during the isolation transition is serializable.
\qed
\end{corollary}

Based on Corollary \ref{cor:transition_ser}, we implement our CIV mechanism by detecting all cross-isolation vulnerable dependencies during the isolation-level transition and ensuring the consistency of the commit and dependency orders. The CIV mechanism includes three steps. (1) When the system transitions from the current isolation $I_{old}$ to the optimal isolation level $I_{new}$, the middle tier blocks new transactions from entering the validation phase until all transactions that have entered the validation phase before the transition commit or abort. Importantly, we only block transactions to enter the validation phase. Transactions can execute normally without blocking. (2) After that, the transaction that has completed the execution phase enters the cross-isolation validation phase. During the cross-isolation validation phase, transactions request validation locks according to the stricter locking method of either $I_{old}$ or $I_{new}$ to ensure that all cross-isolation vulnerable dependencies can be detected. For example, when transitioning from SI to RC, the transaction in the cross-isolation validation phase requests validation locks following RC's validation locking method, regardless of whether it is executed under SI or RC. (3) After acquiring validation locks, transaction $T_i$ first detects vulnerable dependencies of its original isolation level. Then, it detects cross-isolation vulnerable dependencies by checking whether a committed transaction modifies its read set (using the same detection method as that in \S\ref{sec:design:cc:validation}). If such modifications are detected, $T_i$ is aborted to ensure the consistency of the commit and dependency orders. 

Once all transactions executed under $I_{old}$ are committed or aborted, the transition process ends. Then, transactions do not need to undergo the cross-isolation validation.


Specifically, we can prove that the scheduling during the transition between RC and SI is serializable if they perform the concurrency control in \S\ref{design-1} (detailed in \S\ref{sec:proof.switch}).

\section{Serializability and recovery}\label{sec:correctness}
In this section, we first prove the serializability of \sysname's scheduling in the single-isolation level and cross-isolation level categories in \S~\ref{sec:proof.isolation} and \S~\ref{sec:proof.switch}, respectively. Finally, we present the failure recovery strategy in \S~\ref{sec:proof:failure}. 

\subsection{Serializability under Low Isolation Levels \label{sec:proof.isolation}}
Non-serializable scheduling under each low isolation level accommodates certain specific vulnerable dependencies. According to Theorem \ref{the:vulnerable}, a necessary condition for non-serializability is the presence of inconsistent dependencies and commit orders among these vulnerable dependencies. 
\maintext{The unified middle-tier concurrency control ensures the commit order respects dependency order for transactions with vulnerable dependencies, thereby preserving SER even when the RDBMS operates at lower isolation levels.}
\extended{\sysname identifies static vulnerable dependencies from the transaction templates and ensures that, in transactions involving these dependencies, the commit order aligns with the dependency order as specified in Algorithm \ref{alg.transaction}. This approach maintains SER even when the RDBMS is configured to low isolation levels.}

\subsection{Serializability under Cross-isolation Levels \label{sec:proof.switch}}
We prove the correctness of the cross-isolation level in three steps: If there is non-serializable scheduling during the transition, \blackding{1} there exists at least a cross-isolation vulnerable dependency as defined in Definition~\ref{def:transition_vul}; 
\blackding{2} there exists at least a cross-isolation vulnerable dependency $T_j \xrightarrow{rw} T_k$, where $T_k$ commits before $T_j$ and $T_j$ commits after the transition. 
\blackding{3} The cross-isolation validation mechanism can detect the dependency if $T_j$ commits after the transition and enforce the consistent commit and dependency order. \maintext{Due to the space limitation, we provide more details in our technique report \cite{TxnSails}. }




\begin{figure}[t]
    \centering
    \includegraphics[width=0.47\textwidth]{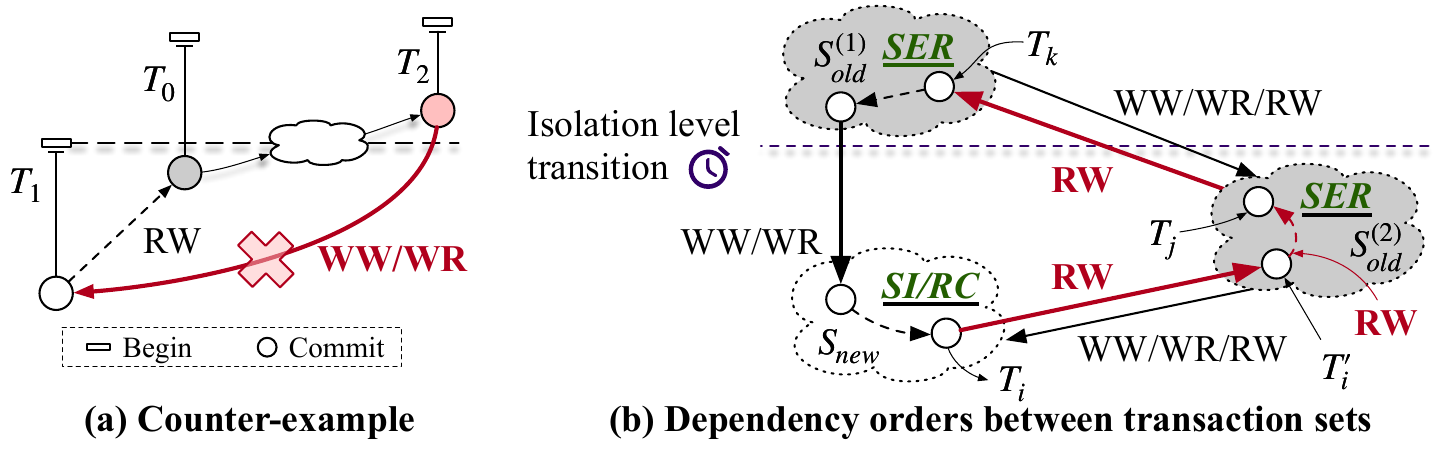}
    \vspace{-4mm}
    \caption{Transition from SER to SI/RC}
    \label{fig:switch_correctness}
    \vspace{-4mm}
\end{figure}

{
\blackding{1} 
\maintext{
We first prove the existence of cross-isolation vulnerable dependency in non-serializable scheduling. 
}
\extended{
We first prove that if there is non-serializable scheduling during the transition, there must be two consecutive RW dependencies, $T_i \xrightarrow{rw} T_j \xrightarrow{rw} T_k$, where $T_k$ commits before $T_j$.
}
If non-serializable scheduling occurs, consider three transactions: $T_2 \xrightarrow{D_1} T_1 \xrightarrow{D_2} T_0$. 
Without loss of generality, we assume $T_0$ is the first transaction committed.
Since $T_0$ commits first, $D_2$ must be an RW dependency; otherwise, $T_1$ should commit before $T_0$. 
Additionally, $T_1$ can not operate under RC because the concurrency control in \S\ref{design-1} would avoid the inconsistent dependency and commit order between $T_1$ and $T_0$. 
Moreover, $D_1$ can only be an RW edge; otherwise, $T_2$ commits before $T_0$ commits, as depicted in Figure~\ref{fig:switch_correctness}a, which contradicts the initial assumption that $T_0$ is the first to commit. 
\extended{Moreover, the data dependency from $T_2$ to $T_1$ can only be an RW edge. We prove this by reductio. If the dependency from $T_2$ to $T_1$ is either a WW or WR dependency, implying that $T_2$ commits before $T_1$ starts. Since $T_1$ is concurrent with $T_0$ due to an RW dependency, deriving $T_0$ commits after $T_1$ starts. Thus, $T_2$ must commit before $T_0$ commits, contradicting the assumption that $T_0$ is the first transaction to commit. Therefore, the data dependency from $T_2$ to $T_1$ must be an RW dependency, leading to two consecutive RW dependencies $T_2 \xrightarrow{rw} T_1 \xrightarrow{rw} T_0$, where $T_0$ commits first. 
}

Moreover, if $T_1$ operates under SI, the concurrency control in \S\ref{design-1} can detect the dependency from $T_1$ to $T_0$ and
enforce the consistent commit and dependency order. 
Hence, if non-serializable scheduling exists, \textbf{$T_1$ must operate under SER}. 

In other words, \textbf{the transition between RC and SI is serializable if they perform the concurrency control in \S\ref{design-1}}. 



\blackding{2} We then prove that the existence of cross-isolation vulnerable dependency $T_j \xrightarrow{rw} T_k$, where $T_k$ commits before $T_j$ and $T_j$ commits after the transition. 
We demonstrate the proof under two cases as follows. 




\noindent\textbf{Transition from SI/RC to SER.} According to proof \blackding{1}, if there is non-serializable scheduling during the transition, there must be two consecutive RW dependencies, $T_i \xrightarrow{rw} T_j \xrightarrow{rw} T_k$, where $T_k$ commits before $T_j$ and $T_j$ operates under SER. During the transition from SI/RC to SER, $T_j$ operates under the new isolation level, making it commit after the transition starts. 

\noindent\textbf{Transition from SER to SI/RC.} For clarity, we categorize the transactions during the transition into three discrete sets:
\maintext{
\begin{itemize}[leftmargin=*]
\item $S_{old}^{(1)}$: Transactions under $I_{old}$ committed before the transition.
\item $S_{old}^{(2)}$: Transactions under $I_{old}$ committed after the transition.
\item $S_{new}$: Transactions under $I_{new}$ starting after the transition.
\end{itemize}
}

\extended{
\begin{itemize}[leftmargin=*]
    \item $S_{old}^{(1)}$: The set of transactions under $I_{old}$ and have been committed when the transition occurs. 
    \item $S_{old}^{(2)}$: The set of transactions that operate under $I_{old}$ and commit after the transition occurs. 
    \item $S_{new}$: The set of transactions that start after the transition occurs and operate under $I_{new}$.
\end{itemize} 
}

Figure \ref{fig:switch_correctness}b shows the partial orders between transaction sets. 
Non-serializable scheduling implies a dependency cycle, 
\maintext{
either (a) between $S_{old}^{(2)}$ and $S_{new}$ or (b) spanning $S_{old}^{(1)}$, $S_{old}^{(2)}$, and $S_{new}$.
}
\extended{
which can be classified into two kinds: (a) scheduling involves only transactions in $S_{old}^{(2)}$ and $S_{new}$; (b) scheduling involves transactions in $S_{old}^{(1)}$, $S_{old}^{(2)}$ and $S_{new}$. 
}

In the first case, all transactions involving vulnerable dependencies are committed after the transition. According to proof \blackding{1}, if there is non-serializable scheduling during the transition, there must be two consecutive RW dependencies, $T_i \xrightarrow{rw} T_j \xrightarrow{rw} T_k$, and $T_j$ commits after the transition. 
In the second case, 
\maintext{
there exists a transaction $T_j\in S_{old}^{(2)}$, its successor transaction is $T_k\in S_{old}^{(1)}$ and its predecessor transaction is either $T_i^{\prime}$ or $T_i$. Hence, there also exists a cross-isolation vulnerable dependency, where $T_k$ commits before $T_j$ and $T_j$ commits after the transition.
} 
\extended{we prove that there is at least one cross-isolation vulnerable dependency, $T_j\xrightarrow{rw} T_k$, where $T_j\in S_{old}^{(2)}$, in the non-serializable scheduling. We conclude that if there is a WR/WW data dependency from $T_i$ to $T_j$, $T_i$ must be committed before $T_i$ starts. Given that, we analyze the possible data dependencies between transaction sets.
The red arrow at the bottom shows that data dependencies from $S_{new}$ to $S_{old}^{(2)}$ can only be RW dependencies due to those transactions in $S_{new}$ commit after those in $S_{old}^{(2)}$.
The red dashed arrow within $S_{old}^{2}$ represents that dependencies between transactions within $S_{old}^{(2)}$ can only be RW dependencies because they are concurrent transactions. 
The red arrow in the top part shows that dependencies from transactions in $S_{old}^{(2)}$ to those in $S_{old}^{(1)}$ must be RW dependencies due to those transactions in $S_{old}^{(2)}$ commit after those in $S_{old}^{(1)}$ start. 
}
\extended{
Next, we use the reductio to prove that transaction $T_j$ in at least one vulnerable dependency, $T_j \xrightarrow{rw} T_k$, is not in the $S_{old}^{(1)}$ set.
If $T_j$ in all cross-isolation vulnerable dependencies, $T_j \xrightarrow{rw} T_k$, is in $S_{old}^{(1)}$, then any transaction $T_j$ in $S_{old}^{(2)}$ which is contained in a RW dependency pointing to $S_{old}^{(1)}$ must not have a precede RW dependency. However, due to that dependencies from transactions in $S_{new}$ (i.e., $T_i$) to $S_{old}^{(2)}$ (i.e., $T_i^\prime$) or dependencies from transactions in $S_{old}^{(2)}$ (i.e., $T_i^\prime$) to $S_{old}^{(2)}$ (i.e., $T_j$) must be RW dependencies, leading to contradiction. Therefore, at least one $T_j$ in cross-isolation vulnerable dependencies, $T_j \xrightarrow{rw} T_k$, is in $S_{old}^{(2)}$ or $S_{new}$. In other words, at least one $T_j$ commits after the transition starts.
}

\blackding{3}
The cross-isolation validation mechanism can detect the vulnerable dependency $T_j \xrightarrow{rw} T_k$ if $T_j$ commits after the transition. It then enforces a consistent commit and dependency order. According to the contrapositive of proof \blackding{2}, if there does not exist cross-isolation vulnerable dependency $T_j \xrightarrow{rw} T_k$, where $T_k$ commits before $T_j$ and $T_j$ commits after the transition, then there is no non-serializable scheduling during the transition. As a result, the scheduling during the transition is serializable. 


}

\subsection{Failure Recovery \label{sec:proof:failure}}
The system incorporates a robust failure recovery mechanism to ensure data consistency and service availability. When \sysname encounters a failure, the system automatically restarts \sysname to re-connect the RDBMS and rolls back all uncommitted transactions. When the RDBMS encounters failures, the system restarts the RDBMS and leverages the ARIES recovery algorithm~\cite{DBLP:journals/tods/MohanHLPS92:ARIES} to recover the database in a consistent state. 

\section{Implementation\label{implementation}}
We implement \sysname from scratch using Java and Python, comprising approximately 6,000 lines of Java code and 500 lines of Python code. \sysname is publicly available via \url{https://github.com/dbiir/TxnSailsServer}. Implemented outside the database kernel, \sysname can seamlessly integrate with any RDBMS that offers the isolation levels defined in~\cite{DBLP:conf/icde/AdyaLO00, DBLP:journals/pvldb/PortsG12}, enhancing performance and ensuring SER.

\begin{table}[t]
    \caption{Interfaces of \sysname}
    \vspace{-4mm}
    \small
    \centering
    \resizebox{\linewidth}{!}{
    \begin{tabular}{ll}
        \Xhline{1.2pt}
        \multicolumn{2}{c}{\textit{Transaction template interfaces}} \\ 
        \Xhline{1.2pt}
        \textit{\makecell[l]{register(template\_name, sql)\\ $\qquad$ $\rightarrow$ SQL ID}}  & \makecell[l]{Register each sql with the template names \\ and receive the sql index in \sysname.} \\ \hline
        \textit{analysis()} & \makecell[l]{Analyze and identify static vulnerable \\ dependencies in low isolation levels.}  \\
        \Xhline{1.2pt}
        \multicolumn{2}{c}{\textit{Transaction execution interfaces}} \\ 
        \Xhline{1.2pt}
        \textit{begin()/commit()/rollback() }   & Begin/Commit/Rollback a transaction. \\ \hline
        \textit{\makecell[l]{execute(template\_name, SQL ID,\\ List[args]) $\qquad$ $\rightarrow$ result}}  & \makecell[l]{Execute the statement with its arguments \\ and receive the execution result.}  \\
        \Xhline{1.2pt}
        \label{tab:api}
    \end{tabular}
    }
    \vspace{-8mm}
\end{table}


\noindent\textbf{Interfaces.} Applications interact with \sysname via predefined interfaces, as detailed in Table~\ref{tab:api}.
To start, applications register transaction templates by \textit{register} interface, which parses SQL templates, extracts operation types and potential data sets, and outputs a unique SQL ID. Then, \textit{analysis} interface is used to identify static vulnerable dependencies under low isolation levels.
During execution, transactions are initiated through \textit{begin} interface, which assigns a unique ID for middle-tier concurrency control. Transactions are executed via \textit{execute} interface, passing the template name, SQL ID, and runtime arguments. \sysname transparently manages concurrency control and isolation levels. Finally, transactions are completed using \textit{commit/rollback} interface.



\noindent\textbf{Analyzer.} We implement \textit{SDGBuilder} class that takes transaction templates as input and constructs a static dependency graph. The graph is then passed to \textit{CycleFinder} class to detect cycles based on the characteristics defined in Theorem~\ref{the:RC}. Finally, it identifies transaction templates with static vulnerable dependencies and stores the results in a \textit{MetaWorker} instance.

\noindent\textbf{Executor.} 
\textit{Executor} invokes \textit{SQLRewrite()} function to rewrite queries, selecting the appropriate record version if its template is involved in static vulnerable dependencies. It then sends the rewritten query to the database and records the \textit{vid} column.
Additionally, we implement a critical data structure, \textit{ValidationMetaTable}, which is initialized before any transactions are received to perform middle-tier validation in both single- and cross-isolation scenarios. Organized as a hash table, each bucket in the table represents a list of \textit{ValidationMeta} entries, including \textit{validation lock}, \textit{latest version}, and \textit{lease} information.
A dedicated thread handles garbage collection of expired meta entries by comparing the \textit{lease} with the system's real-time clock. Moreover, we implement the \textit{WAIT-DIE} strategy within the \textit{ValidationMetaTable} to prevent deadlocks.

\noindent\textbf{Adapter.} 
We first implement \textit{TransactionCollector} class that collects the read and write sets for transactions. Then, we implement a {\it RDGBuilder} class to build the runtime dependency graph. Finally, {\it Adapter} is implemented with the aid of \textit{torch\_geometric}, taking the runtime dependency graph as input and outputting the optimal isolation level.
To ensure cross-platform compatibility and efficiency, the Python and Java components communicate via \textit{sockets}. 

\section{evaluations\label{sec:evaluation}}

In this section, we evaluate \sysname's performance compared to state-of-the-art solutions. 
Our goal is to validate two critical aspects empirically: (1) \sysname's effectiveness in adaptively selecting the appropriate isolation level for dynamic workloads (\S\ref{sec:evaluation:overall}); and (2) \sysname's performance superiority over state-of-the-art solutions across a variety of scenarios (\S\ref{sec:evaluation:compare-to-other-soluation}). 

\subsection{Setup}
We run our database and clients on two separate in-cluster servers with an Intel(R) Xeon(R) Platinum 8361HC CPU @ 2.60GHz processor, which includes 24 physical cores, 64 GB DRAM, and 500 GB SSD. 
The operating system is CentOS Linux release 7.9. 


\subsubsection{Default configuration.}
We utilize BenchBase~\cite{DBLP:journals/pvldb/DifallahPCC13} as our benchmark simulator, 
which is deployed on the client server. By default, the experiments are conducted using 128 client terminals.
We deployed PostgreSQL 15.2~\cite{PostgreSQL} as the database engine, which employs MVCC to implement three distinct isolation levels: Read Committed (RC), Snapshot isolation (SI), and Serializable (SER) (by SSI~\cite{DBLP:conf/sigmod/CahillRF08}). 
Under RC, the system can read the most recently committed version, while both SI and SER maintain a view of the data as it existed at the start of the transaction, thereby observing the committed version from that point in time. To prevent dirty writes, write locks are enforced at all isolation levels.
For our database configuration, we allocated a buffer pool size of 24GB, limited the maximum number of connections to 2000, and established a lock wait timeout of 100 ms. 
To eliminate network-related effects, both \sysname and PostgreSQL were deployed on another server.

\subsubsection{Baselines.} To ensure a fair comparison, we implemented existing approaches within the BenchBase framework and connected them directly to PostgreSQL.  

\noindent\textbf{Baselines within the database kernel.} 
We evaluated concurrency control algorithms supported natively by PostgreSQL, specifically those associated with lower isolation levels that can achieve serializable scheduling: 

\textit{(1) \& (2) PostgreSQL’s native concurrency control mechanisms (\textbf{SER} and \textbf{SI}).}  
These approaches execute workloads configured at the SER or SI levels without requiring workload modifications. For instance, TPC-C achieves serializable scheduling under SI, while SmallBank requires SER for serializability. 



\noindent\textbf{Baselines outside the database kernel.} We also evaluated external strategies that transform RW dependencies into WW dependencies to eliminate static dangerous structures.  


\textit{(3) \& (4) Promotion (\textbf{RC+Promotion}\cite{DBLP:conf/icdt/VandevoortK0N22}, \textbf{SI+Promotion} \cite{DBLP:conf/icde/AlomariCFR08}).} 
This strategy converts read operations into write operations by promoting \textit{SELECT} statements with non-modifying \textit{UPDATE} statements~\cite{DBLP:conf/icde/AlomariCFR08}. These modifications are applied at RC and SI levels, referred to as RC+Promotion and SI+Promotion, respectively.  


\textit{(5) \& (6) Conflict materialization (\textbf{RC+ELM}~\cite{DBLP:conf/aiccsa/AlomariF15}, \textbf{SI+ELM} \cite{DBLP:conf/icde/AlomariCFR08}).} 
This approach employs an external lock manager (ELM) and introduces additional write operations on the ELM to ensure serializable scheduling. It is applied at both RC and SI levels, referred to as RC+ELM and SI+ELM, respectively.

For each strategy, we adopted the most effective variant as identified in prior work~\cite{DBLP:conf/aiccsa/AlomariF15}. For example, under the Promotion strategy in the SmallBank benchmark with SI, modifying the \textit{WriteCheck} template rather than the \textit{Balance} template yielded superior performance.  

Finally, we evaluate the middle-tier concurrency control in \S\ref{design-1} at both RC and SI levels without self-adaptive isolation level selection, denoted as \textbf{\sysname-RC} and \textbf{\sysname-SI}, respectively.

\subsubsection{Benchmarks.} Three benchmarks are conducted as follows. 

\noindent\textbf{SmallBank~\cite{DBLP:conf/icde/AlomariCFR08}.} This benchmark populates the database with 400k accounts, each having associated checking and savings accounts. 
Transactions are selected by each client using a uniform distribution. To simulate transactional access skew, we employ a Zipfian distribution with a default \textit{skew factor} of 0.7.

\noindent\textbf{YCSB+T~\cite{DBLP:conf/icde/DeyFNR14}.} 
This benchmark generates synthetic workloads emulating large-scale Internet applications. In our setup, the \textit{usertable} consists of 10 million records, each 1KB in size, totaling 10GB. The \textit{skew factor}, set by default to 0.7, controls the distribution of accessed data items, with higher values increasing data contention. Each default transaction involves 10 operations, with a 90\% probability of being a read and a 10\% probability of being a write.

\noindent\textbf{TPC-C~\cite{TPCC}.} We use the TPC-C benchmark, which modifies the schema and templates to convert all predicate reads into key-based accesses according to our baseline~\cite{DBLP:journals/pvldb/VandevoortK0N21}. It includes 5 transaction templates: NewOrder, Payment, OrderStatus, Delivery, and StockLevel. Our tests host 32 warehouses, with each containing about 100MB of data. Following previous works~\cite{DBLP:journals/pvldb/YuBPDS14, DBLP:journals/pvldb/HardingAPS17}, we exclude user data errors that cause about 1\% of NewOrder transactions to abort. 


\subsection{Ablation Study}
\label{sec:evaluation:overall}

\begin{figure}[t]
    \centering
    \begin{minipage}{0.8\linewidth}
        \centering
        \includegraphics[width=\linewidth]{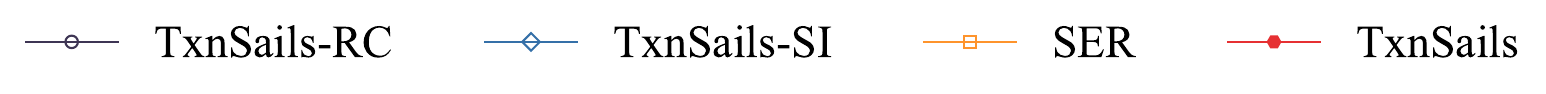}
        \vspace{-5mm}
    \end{minipage}
    \begin{minipage}{0.95\linewidth}
        \centering
        \begin{subfigure}{0.95\linewidth}
            \includegraphics[width=\linewidth]{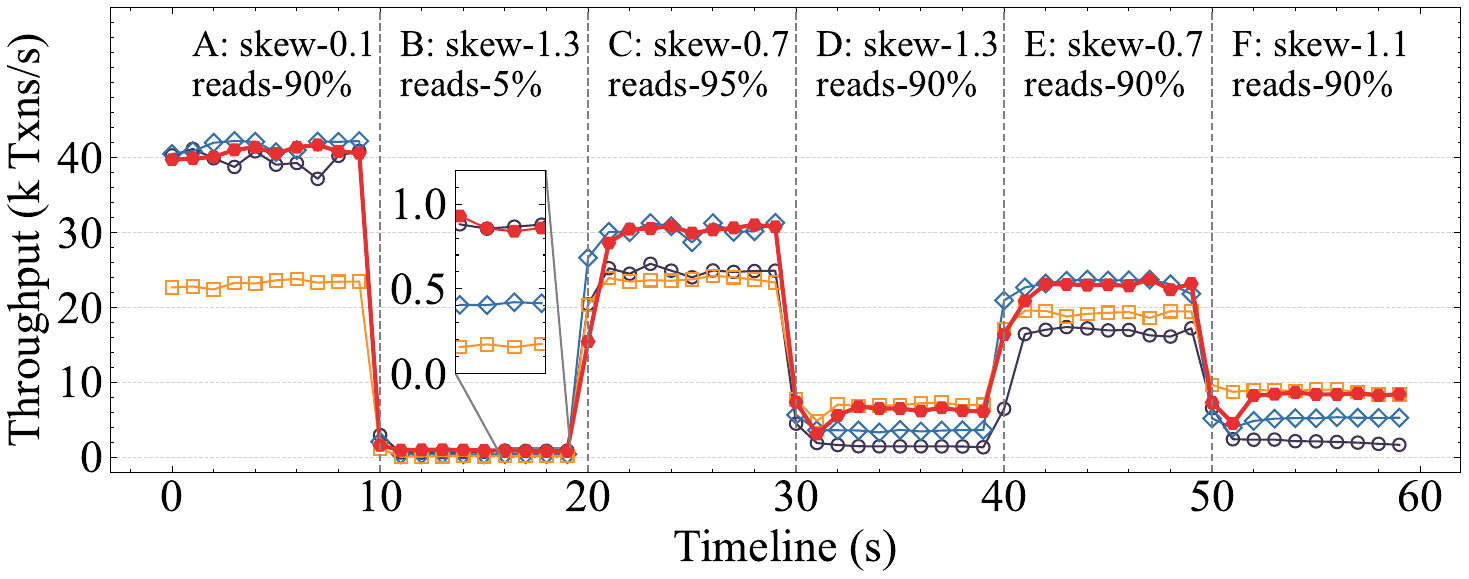}
            \vspace{-4mm}
        \end{subfigure}
    \end{minipage}
    \vspace{-4mm}
    \caption{Workload shifting by YCSB}
    \label{fig:evaluation.dynamic}
    \vspace{-6mm}
\end{figure}

In this part, we evaluate the effectiveness of the self-adaptive isolation level selection and isolation transition in \sysname.

\subsubsection{Self-adaptive isolation level selection}
We first evaluate the selection of self-adaptive isolation level by varying the workload every 10 seconds across six distinct scenarios. The experimental results are illustrated in Figure~\ref{fig:evaluation.dynamic}. 
We sample the workload at 1-second intervals. 
The results demonstrate that different isolation levels perform variably under different workloads: SI performs well in low-skew scenarios (A, C, E), SER is more effective in high-skew scenarios with a lower percentage of write operations (D, F), and RC excels in high skew scenarios with a high percentage of writes (B). Across all tested dynamic scenarios, \sysname successfully adapts to optimal isolation level. Specifically, the graph learning model in \sysname identifies that SI is suitable for scenarios with fewer conflicts due to its higher concurrency and lower data access overhead (i.e., one-time timestamp acquisition). Conversely, RC is ideal for scenarios with higher conflict rates and more write operations, as it efficiently handles concurrent writes (SI aborts concurrent writes, while RC allows them to commit). Compared to an application directly run on RDBMS at the SER level, the performance of \sysname when selecting SER is slightly reduced by 4.3\%. This overhead arises from two aspects: (1) \sysname requires modifications to the application code to select \textit{version} and another localhost message delivery ; (2) \sysname needs to sample transactions to predict the optimal isolation level, even though this is an asynchronous task.

\subsubsection{Validation analysis\label{sec:evaluation:transition}}
This part evaluates the validation efficiency in both single-isolation and cross-isolation scenarios. 

\begin{figure}[t]
    \centering
    \begin{subfigure}{0.47\linewidth}
        \centering
        \includegraphics[width=\linewidth]{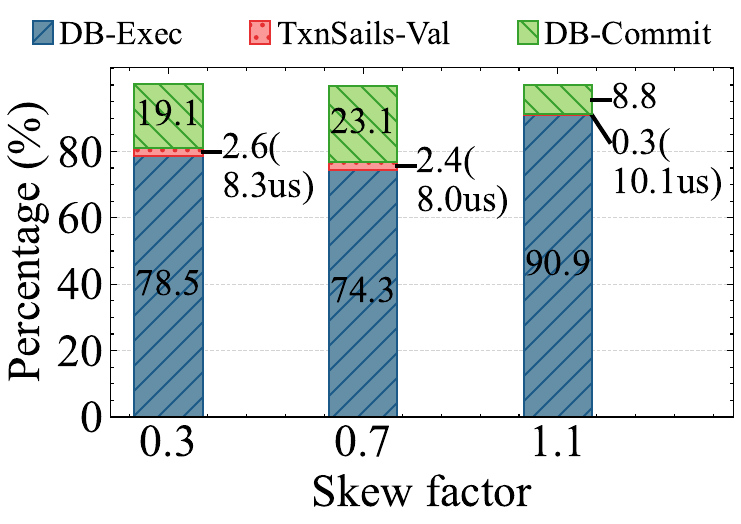}
        \vspace{-6mm}
        \caption{Transaction breakdown}
        \label{fig:evaluation.breakdown.transaction}
    \end{subfigure}
    \begin{subfigure}{0.47\linewidth}
        \centering
        \includegraphics[width=\linewidth]{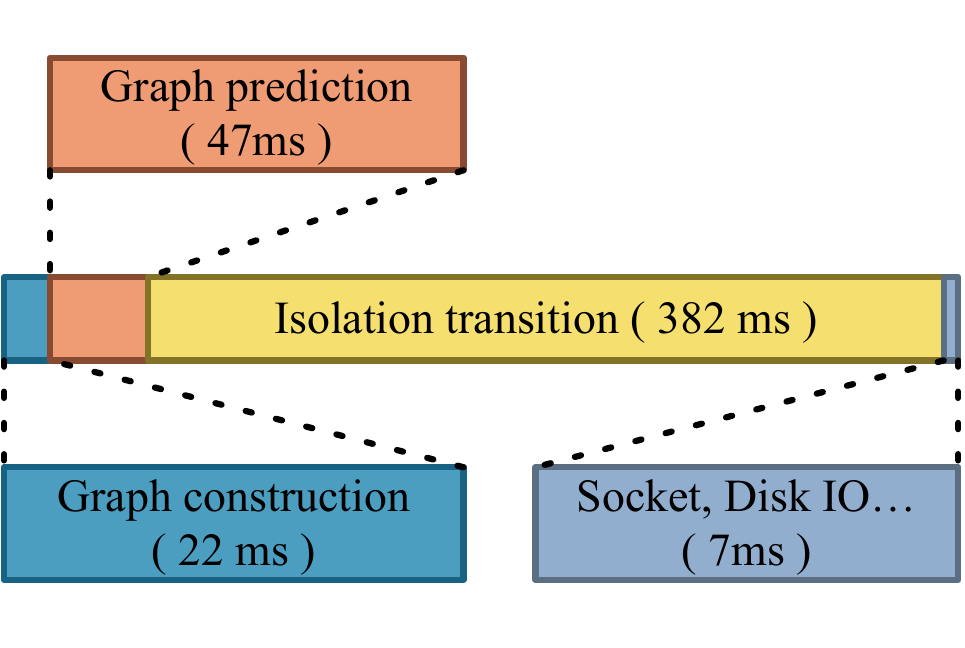}
        \vspace{-6mm}
        \caption{Transition breakdown}
        \label{fig:evaluation.breakdown.transition}
    \end{subfigure}
    \vspace{-4mm}
    \caption{Breakdown analysis by YCSB}
    \label{fig:evaluation.dynamic.breakdown}
    \vspace{-4mm}
\end{figure}

\noindent\textbf{Single-isolation level validation.} 
We evaluate the single-level validation cost under skew factors of 0.3, 0.7, and 1.1, respectively. The average breakdown is depicted in Figure~\ref{fig:evaluation.breakdown.transaction}. The validation cost remains relatively stable, decreasing from 2.6\% to 0.3\% of the transaction lifecycle as contention increases. This suggests that the middle-tier concurrency control proposed in \S\ref{design-1} does not significantly affect normal execution.

\begin{figure}[t]
    \centering
    \begin{minipage}{0.95\linewidth}
        \centering
        \includegraphics[width=\linewidth]{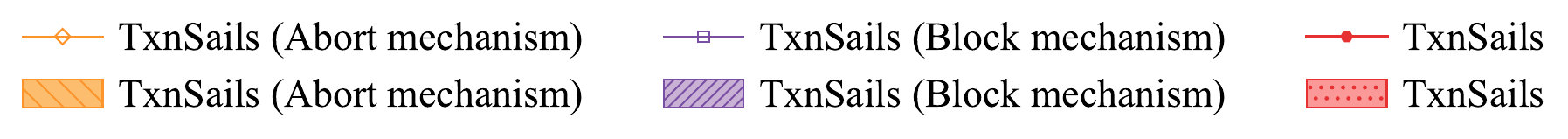}
        \vspace{-5mm}
    \end{minipage}
    \begin{minipage}{0.95\linewidth}
        \centering
        \begin{subfigure}{0.48\linewidth}
            \includegraphics[width=\linewidth]{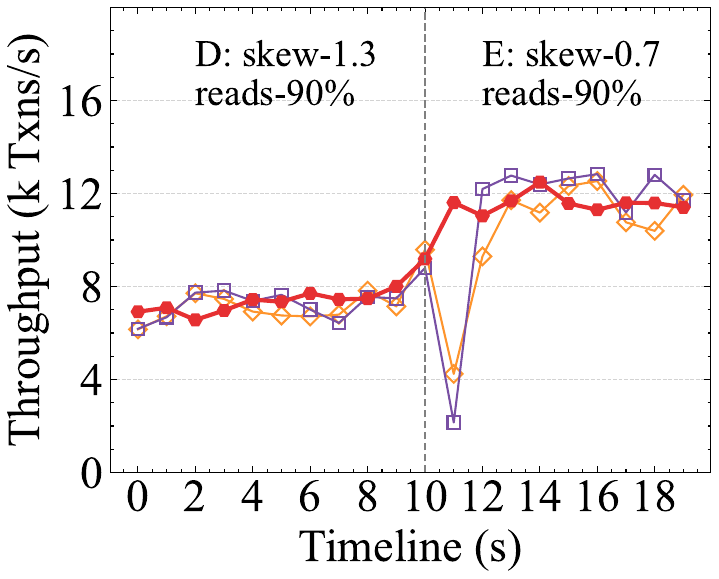}
            \vspace{-6mm}
            \caption{Transition from D to E}
            \label{fig:evaluation.dynamic_switch.de}
            \vspace{-4mm}
        \end{subfigure}
        \begin{subfigure}{0.48\linewidth}
            \includegraphics[width=\linewidth]{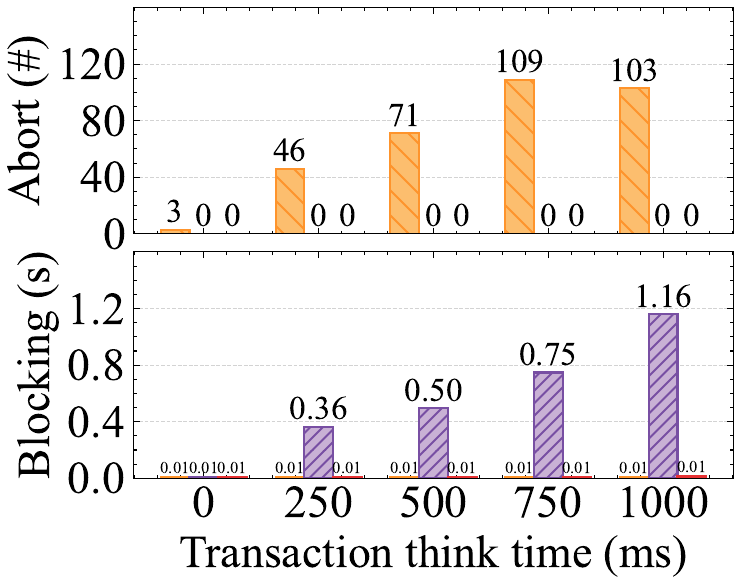}
            \vspace{-6mm}
            \caption{Performance metrics}
            \label{fig:evaluation.dynamic_switch.metric}
            \vspace{-4mm}
        \end{subfigure}
    \end{minipage}
    \caption{Comparasion of transition mechanisms by YCSB}
    \label{fig:evaluation.dynamic_switch}
    \vspace{-6mm}
\end{figure}

\noindent\textbf{Cross-isolation level validation.} 
We evaluate \sysname with YCSB using the various transition mechanisms mentioned in \S\ref{design-3}. We first evaluate the transition of the workload from D to E with a ``think time'' of 1s, as illustrated in Figure~\ref{fig:evaluation.dynamic_switch.de}. \sysname minimizes the impact of isolation level transitions by avoiding active block time or aborts, maintaining serializable scheduling.
To further compare mechanisms, we vary the ``think time'' parameter (Figure~\ref{fig:evaluation.dynamic_switch.metric}). Increased think time raises transaction latency and leads to more aborts under the abort strategy, while the blocking strategy incurs longer blocking times. In contrast, the cross-isolation validation mechanism outperforms both, reducing transaction aborts and blocking time while delivering performance improvements of up to 2.7$\times$ and 5.4$\times$, respectively.


\subsubsection{Graph model: construction, training, and prediction}

Figure~\ref{fig:evaluation.breakdown.transition} illustrates the overhead of workload transition, which takes approximately 450 milliseconds.
Specifically, graph construction and prediction require 22 milliseconds and 47 milliseconds, respectively, while over 80\% of the time is spent on transition, from initiating the transition to all connections adopting the new isolation level, closely tied to the longest transaction execution latency.
Notably, the prediction in Figure \ref{fig:evaluation.dynamic} is inaccurate for 1 or 2 seconds at the 30-second and 50-second marks due to the sampling transactions from the previous workload during the transition. However, the model successfully transitions to the optimal isolation level in subsequent prediction cycles. The overhead of the learned model is minimal, with less than a 2.5\% difference in throughput between using the graph model and not using it.

\begin{figure}[t]
    \centering
    \begin{minipage}{0.95\linewidth}
        \centering
        \begin{subfigure}{0.48\linewidth}
            \includegraphics[width=\linewidth]{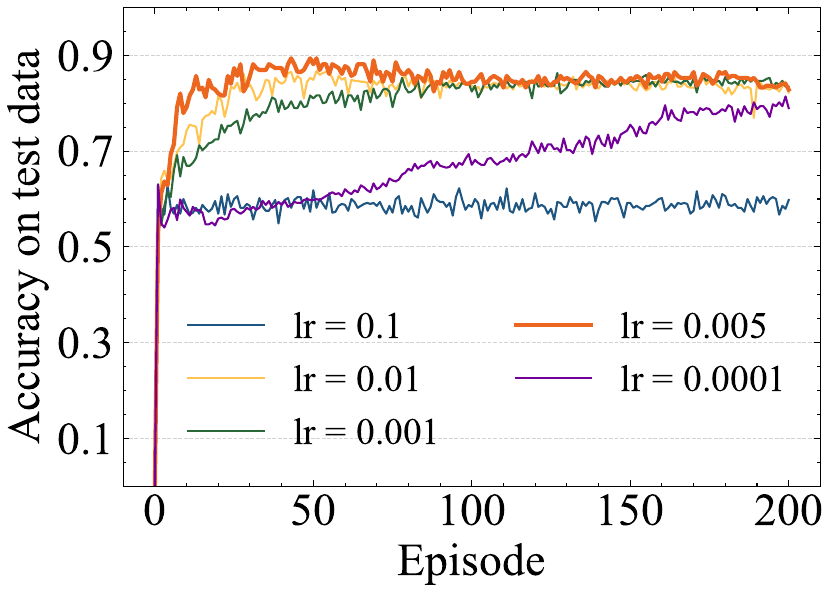}
            \vspace{-6mm}
            \caption{Accuarcy}
            \label{fig:evaluation.train.acc}
        \end{subfigure}
        \begin{subfigure}{0.48\linewidth}
        \includegraphics[width=\linewidth]{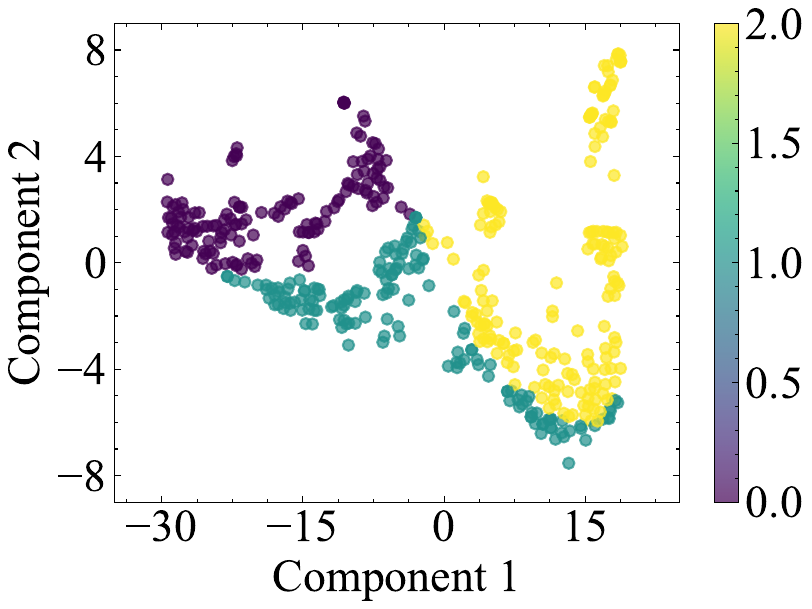}
            \vspace{-6mm}
            \caption{Extracted features}
            \label{fig:evaluation.train.cluster}
        \end{subfigure}
    \end{minipage}
    \vspace{-4mm}
    \caption{Model training metrics by YCSB}
    \label{fig:evaluation.train}
    \vspace{-4mm}
\end{figure}

\extended{
\begin{figure}[t]
    \centering
    \begin{minipage}{0.95\linewidth}
        \centering
        \begin{subfigure}{0.48\linewidth}
            \includegraphics[width=\linewidth]{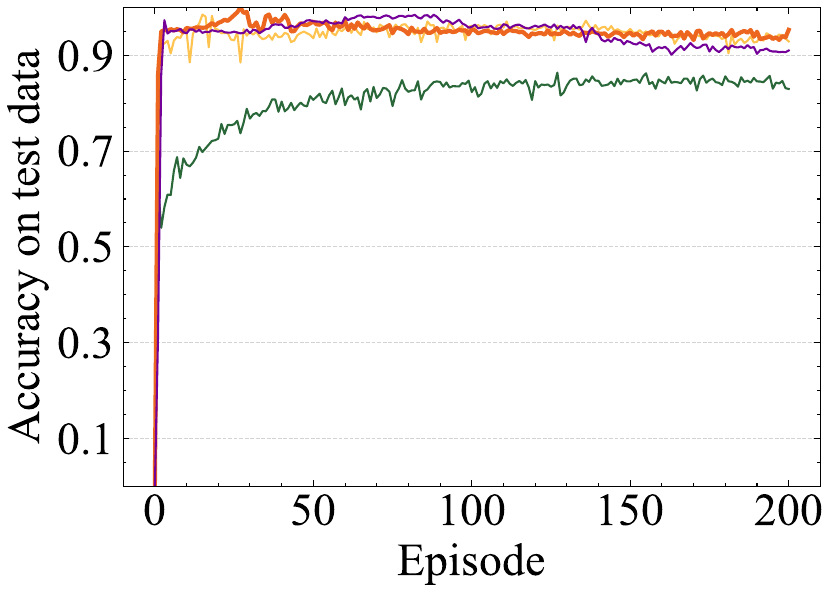}
            \vspace{-4mm}
            \caption{Accuarcy by smallbank}
            \label{fig:evaluation.train.sb.acc}
        \end{subfigure}
        \begin{subfigure}{0.48\linewidth}
        \includegraphics[width=\linewidth]{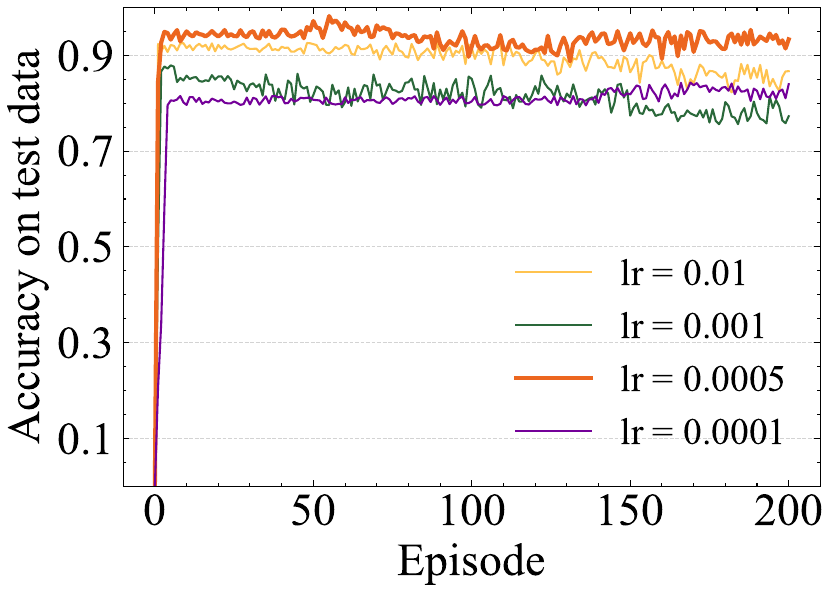}
            \vspace{-4mm}
            \caption{Accuarcy by TPC-C}
            \label{fig:evaluation.train.tpcc.acc}
        \end{subfigure}
    \end{minipage}
    \vspace{-4mm}
    \caption{Model training metrics}
    \label{fig:evaluation.train.tpcc_sb}
    \vspace{-6mm}
\end{figure}
}

We also compare the training process at various learning rates. As shown in Figure~\ref{fig:evaluation.train.acc}, we find that a learning rate of 0.005 quickly achieves approximately 86\% accuracy on test workloads. A small learning rate (0.0001) results in slow training due to minimal weight updates, while a large learning rate (0.1 or greater) can lead to poor accuracy.
To visualize the high-dimensional vectors produced by our model, we use t-SNE~\cite{tsne} for nonlinear dimensionality reduction, mapping them into two dimensions and plotting them with their true labels in Figure~\ref{fig:evaluation.train.cluster}. Most workloads are accurately distinguished, with errors primarily occurring at the boundaries between isolation levels, where performance similarities can lead to incorrect predictions that do not significantly impact overall performance.
\maintext{For more detailed descriptions of the other two models, please refer to our technical report~\cite{TxnSails}.}
\extended{
We further illustrate the training accuracy for Smallbank and TPC-C in Figure~\ref{fig:evaluation.train.tpcc_sb}. During the training process, we noticed an imbalance among the three types of labels. For example, in the Smallbank benchmark, TxnSails-SI consistently outperformed the other two in various scenarios. Inspired by downsampling techniques, we reduce the training data for certain classifications to balance the dataset. We set the model's learning rate to 0.0005. After 10 rounds of training, the accuracy stabilizes at 95.1\% for Smallbank and 91.7\% for TPC-C.
}

\noindent \textbf{\underline{Summary.}} One isolation level does not fits all workloads. 
In low-skew scenarios, SI outperforms RC; in high-skew scenarios with fewer writes, SER is the most effective; and in high-skew scenarios with intensive writes, RC is more suitable. 
\sysname effectively guarantees SER at lower levels and efficiently adapts isolation levels to optimize performance for dynamic workloads using the proposed fast isolation level transition technique.


\subsection{Comparision to State-of-the-art Solutions}
\label{sec:evaluation:compare-to-other-soluation}

We evaluate \sysname against state-of-the-art solutions that use \textbf{external lock manager (ELM)} \cite{DBLP:conf/icde/AlomariCFR08,DBLP:conf/aiccsa/AlomariF15} and \textbf{Promotion} \cite{DBLP:conf/icde/AlomariCFR08,DBLP:conf/icdt/VandevoortK0N22}  over workloads by YCSB, SmallBank, and TPC-C benchmarks.  

\subsubsection{Impact of data contention}
\label{sec:evaluation:contention}

This part studies the impact of data contention by varying the \textit{skew factor} (SF) 
and by varying \textit{hotspot probability} 
and the number of hotspots to simulate different data contention in YCSB and SmallBank, respectively.

In YCSB, \sysname outperforms other solutions by up to 22.7$\times$ and is 2.4$\times$ better than the second-best solution due to lightweight validation without workload modification, thus higher concurrency as depicted in Figure~\ref{fig:ycsb.contention}. In this case, SOTA solutions can not beat SER as they introduce additional write operations in YCSB workloads that restrict concurrency. In high contention scenarios (SF>0.9), validation costs outweigh the benefits of using a lower isolation level, triggering \sysname to transition to the SER level and perform slightly (<5\%) lower than SER due to \sysname overhead. 
In low contention scenarios (SF<0.9), the ELM approaches yield better efficiency than the Promotion ones as lock conflicts are low with external locks. 
We further analyze the latency distribution with the skew factor of 0.9 using cumulative distribution function (CDF) plots, as shown in Figure~\ref{fig:ycsb.skew.cdf.09}. In all scenarios, \sysname reduces the latency of transactions. 

\begin{figure}[]
    \centering
    \begin{minipage}{0.8\linewidth}
        \centering
        \includegraphics[width=\linewidth]{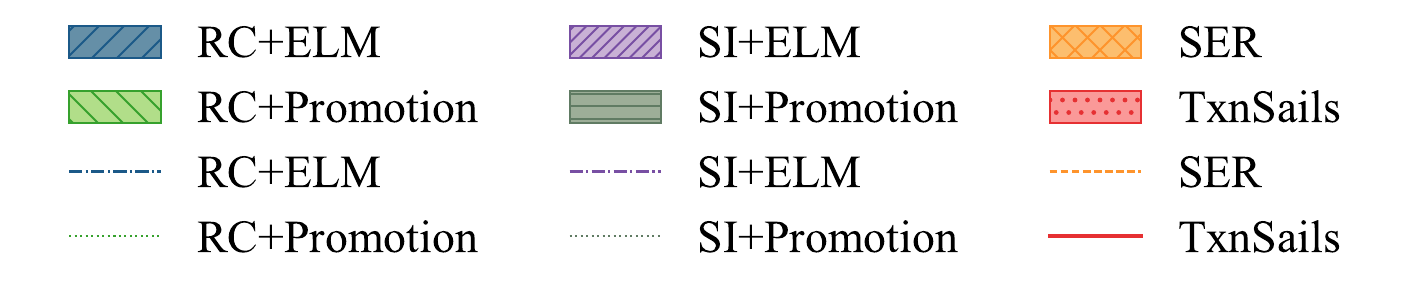}
        \vspace{-5mm}
    \end{minipage}
    \begin{minipage}{0.95\linewidth}
        \centering
        \begin{subfigure}{0.46\linewidth}
            \includegraphics[width=\linewidth]{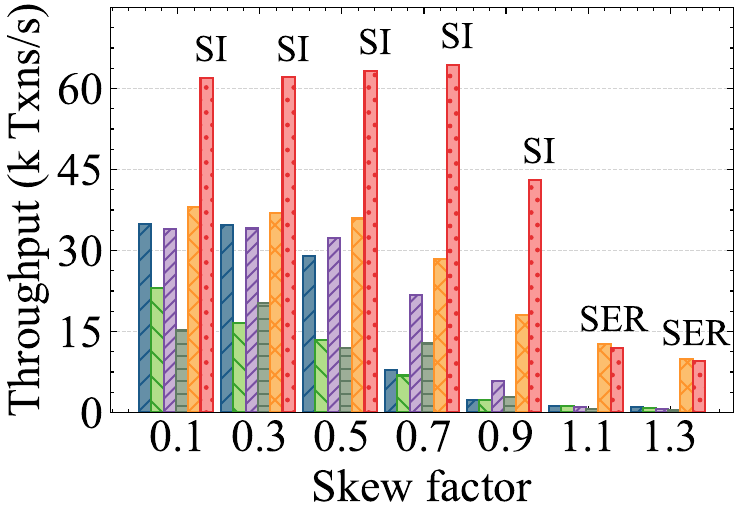}
            \vspace{-6mm}
            \caption{Performance}
            \label{fig:ycsb.contention}
        \end{subfigure}
        \begin{subfigure}{0.46\linewidth}
            \includegraphics[width=\linewidth]{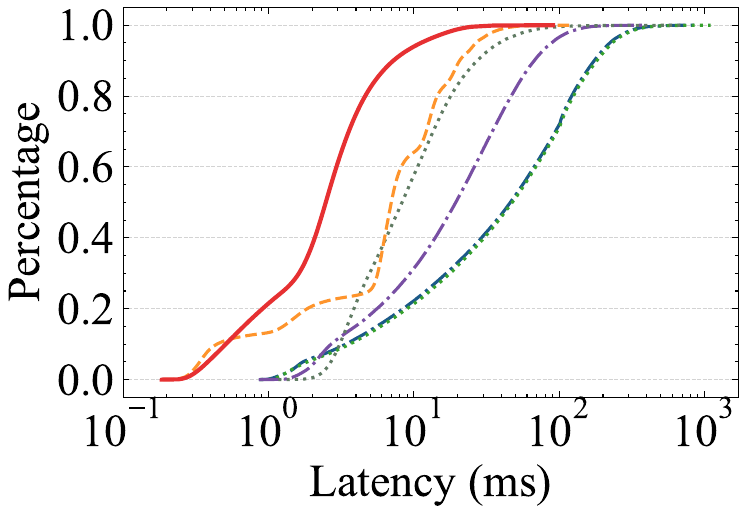}
            \vspace{-6mm}
            \caption{Analysis of latency CDF}
            \label{fig:ycsb.skew.cdf.09}
        \end{subfigure}
    \end{minipage}

    \vspace{-4mm}
    \caption{Impact of data contention by YCSB}
    \label{fig:evaluation.contention.ycsb.cdf}
    \vspace{-4mm}
\end{figure}

In SmallBank, \sysname consistently outperforms other solutions by up to 15.27$\times$ improvement and 2.06$\times$ better than the second-best solution, as depicted in Figure \ref{fig:evaluation.contention.sb}. This time, SI+Promotion can outperform SER. The reason is that, unlike YCSB, SmallBank validates only a small portion of read and write operations. As the skew factor increases, \sysname maintains its advantage over SER by employing SI level. 
As hotspot size increases, the reduced conflicts between transactions make the performance advantage of \sysname less pronounced. 


\begin{figure}[]
    \centering
    \begin{minipage}{0.8\linewidth}
        \centering
        \includegraphics[width=\linewidth]{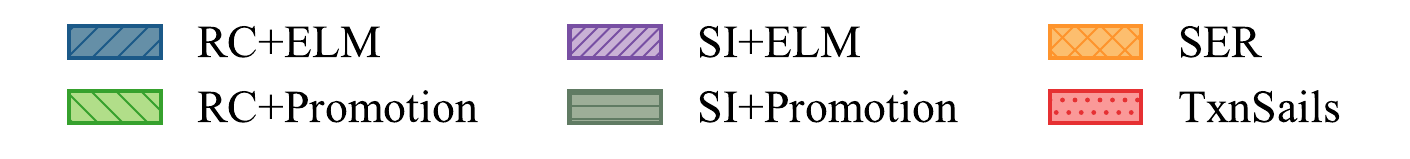}
        \vspace{-5mm}
    \end{minipage}
    \begin{minipage}{0.95\linewidth}
        \centering
        \begin{subfigure}{1.0\linewidth}
            \includegraphics[width=\linewidth]{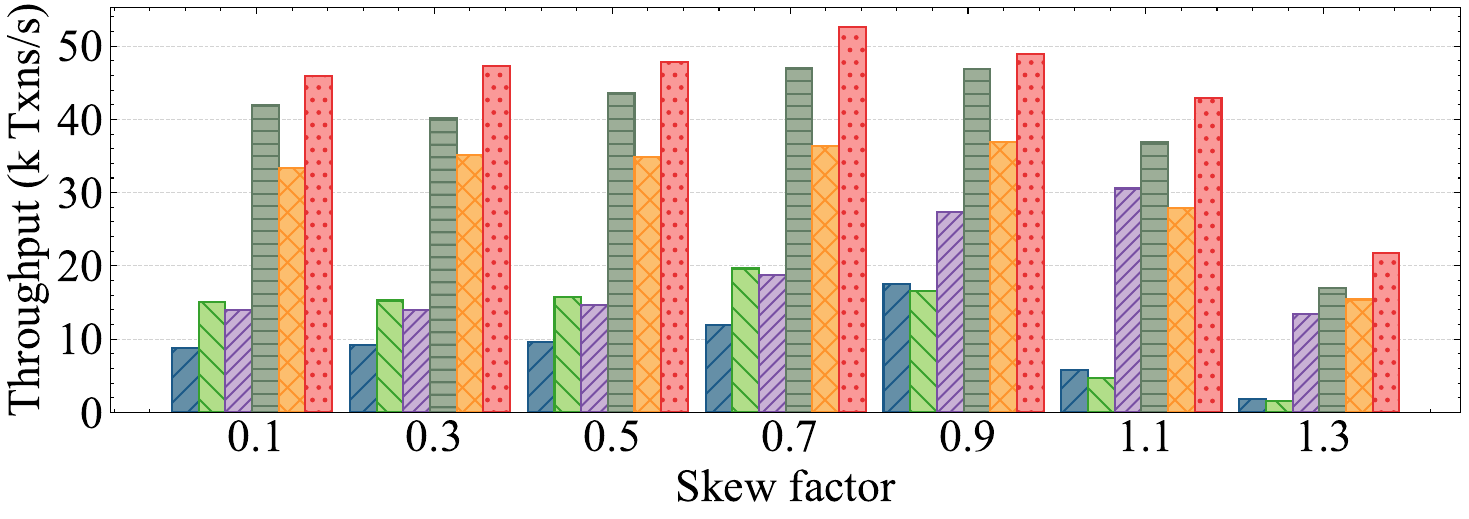}
            \vspace{-6mm}
            \caption{Performance with \textit{skew factors}}
            \label{fig:sb.skew.per}
        \end{subfigure}
        \begin{subfigure}{1.0\linewidth}
            \includegraphics[width=\linewidth]{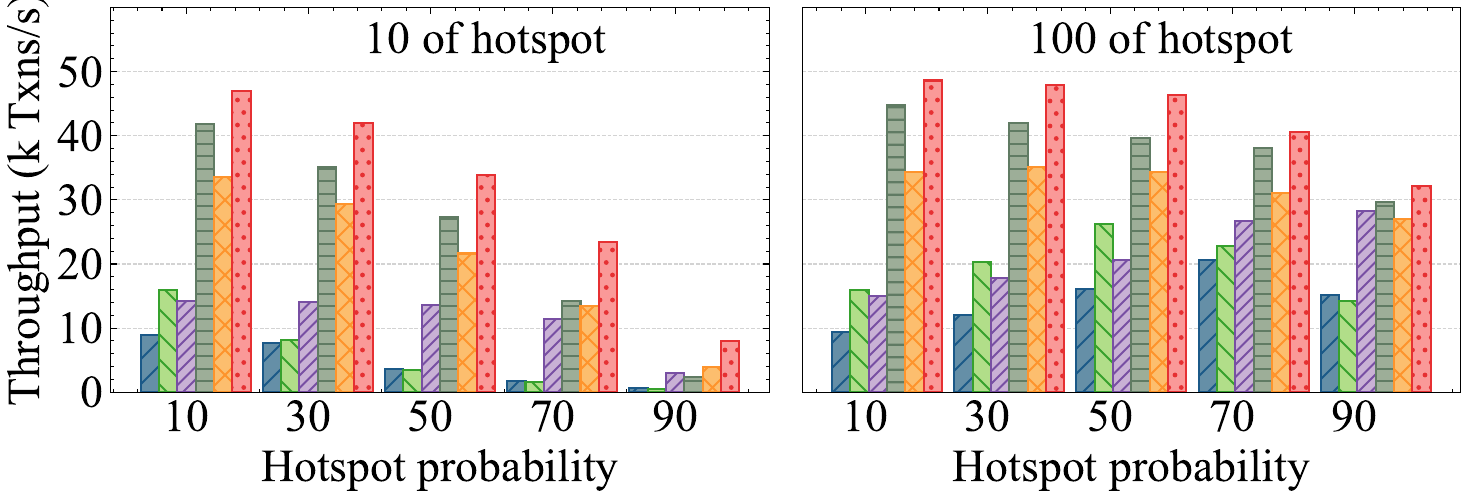}
            \vspace{-6mm}
            \caption{Performance with fixed number of hotspots}
            \label{fig:sb.hotspot.per}
        \end{subfigure}
    \end{minipage}
    \vspace{-4mm}
    \caption{Impact of data contention by SmallBank}
    \label{fig:evaluation.contention.sb}
    \vspace{-6mm}
\end{figure}

\subsubsection{Impact of write/read ratios}
\label{sec:evaluation:wr}
This part evaluates the performance of varying the percentage of write operations with YCSB, using the \textit{skew factors} of 0.1 and 0.7. 
In read-write scenarios in Figure~\ref{fig:ycsb.wr.skew01} and~\ref{fig:ycsb.wr.skew07}, \sysname can outperform other solutions up to 6.68x. As the percentage of write operations increases, the performance gap narrows as verification overhead at lower isolation levels increases. \sysname transitions from using SI to SER and finally to RC, as the FCW \cite{DBLP:journals/pvldb/ChenPLYHTLCZD24_TDSQL} strategy increases the abort rate in scenarios with a high percentage of write operations.

We also evaluate the performance in read-only scenarios in Figure~\ref{fig:ycsb.wr.ro}. \sysname achieves performance up to 4.6$\times$ higher than SER and up to 20.4$\times$ higher than others. \sysname adopts to SI level as its in-memory validation is nearly costless and rarely fails. Other solutions convert read operations to write operations, thereby restricting concurrency. 
This also highlights that when a database is configured to be SER, there is a significant performance loss compared to SI, even with read-only scenarios. 

\begin{figure}[]
    \centering
    \begin{minipage}{0.8\linewidth}
        \centering
        \includegraphics[width=\linewidth]{figures/evaluation/bar_legend01.pdf}
        \vspace{-5mm}
    \end{minipage}
    \begin{minipage}{0.95\linewidth}
        \centering
        \begin{subfigure}{0.48\linewidth}
            \includegraphics[width=\linewidth]{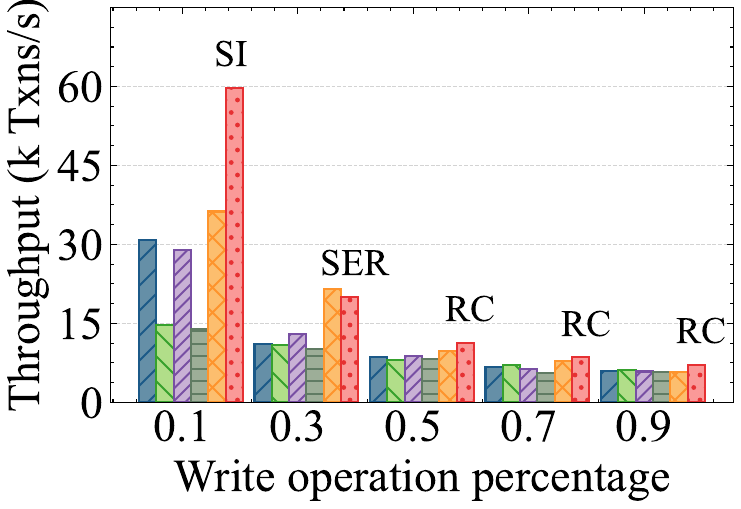}
            \vspace{-6mm}
            \caption{Skew factor is 0.1 - YCSB}
            \label{fig:ycsb.wr.skew01}
        \end{subfigure}
        \begin{subfigure}{0.48\linewidth}
            \includegraphics[width=\linewidth]{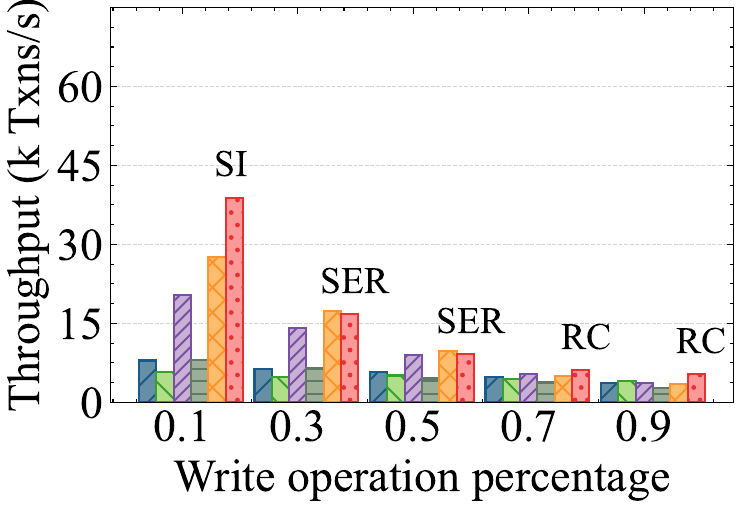}
            \vspace{-6mm}
            \caption{Skew factor is 0.7 - YCSB}
            \label{fig:ycsb.wr.skew07}
        \end{subfigure}
        \begin{subfigure}{0.9\linewidth}
            \vspace{1mm}   
            \includegraphics[width=\linewidth]{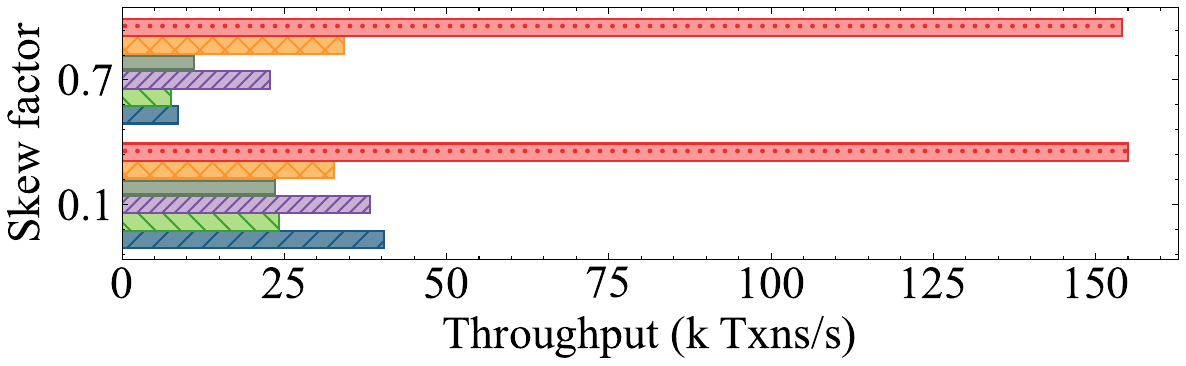}
            \vspace{-6mm}
            \caption{Read only transactions - YCSB}
            \label{fig:ycsb.wr.ro}
        \end{subfigure}
    \end{minipage}
    \vspace{-4mm}
    \caption{Impact of write/read ratio by YCSB}
    \label{fig:evaluation.wr}
    \vspace{-4mm}
\end{figure}

\subsubsection{Impact of templates percentages}
\label{sec:evaluation:ratio}
In complex workloads like SmallBank and TPC-C, only certain transaction templates lead to data anomalies, so modifying these templates can ensure serializability under low isolation levels. This part compares different solutions by varying the percentage of critical transaction templates. 

\begin{figure}[]
    \centering
    \begin{minipage}{0.8\linewidth}
        \centering
        \includegraphics[width=\linewidth]{figures/evaluation/bar_legend01.pdf}
        \vspace{-5mm}
    \end{minipage}
    \begin{minipage}{0.95\linewidth}
        \centering
        \begin{subfigure}{0.95\linewidth}
            \includegraphics[width=\linewidth]{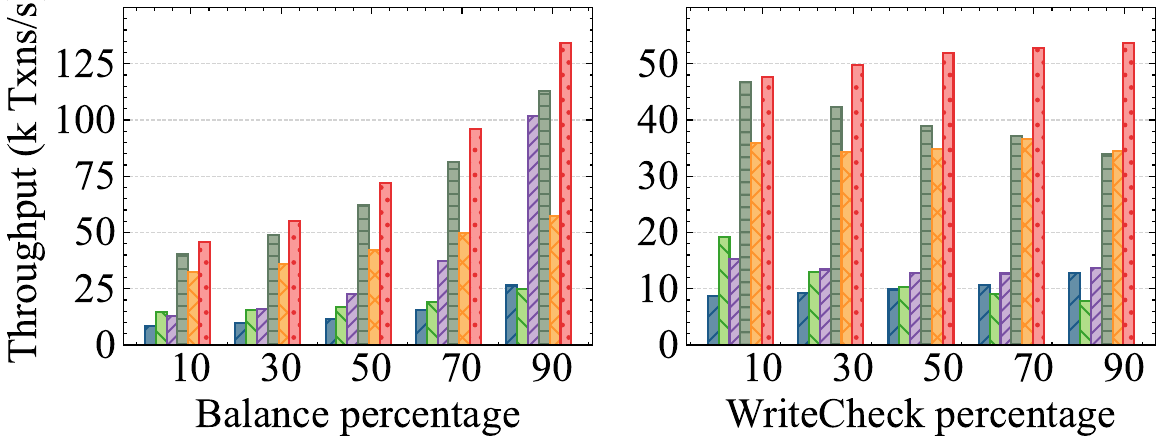}
        \end{subfigure}
    \end{minipage}
    \vspace{-4mm}
    \caption{Impact of templates percentage by SmallBank}
    \label{fig:evaluation.sb.ratio}
    \vspace{-6mm}
\end{figure}

In SmallBank, we evaluate the proportions of the read-only \textit{Balance} and write transaction \textit{WriteCheck}, as shown in Figure~\ref{fig:evaluation.sb.ratio}. 
As the proportion of \textit{Balance} transactions increases, performance improves; however, RC+ELM and RC+Promotion introduce additional writes in \textit{Balance}, leading to increased WW conflicts. In contrast, SI+ELM and SI+Promotion perform better since they do not modify read-only \textit{Balance} transactions.
\sysname-RC must detect RW dependencies from \textit{Balance}, increasing overhead as their proportion rises. Thus, \sysname transitions to SI in this scenario, achieving up to a 6.2$\times$ performance gain over RC+ELM and RC+Promotion. Conversely, as the proportion of \textit{WriteCheck} transactions increases, concurrency decreases, leading to worse performance for SI+ELM and SI+Promotion. However, \sysname's performance advantage becomes more pronounced as it maintains consistent commit and dependency orders through validation without modifying the workload. At 90\% \textit{WriteCheck} transactions, \sysname improves performance by 58.1\% compared to SI+Promotion and achieves 2.3$\times$ the performance of SER.

TPC-C can execute serializable under SI, eliminating the need for validation in SI. The critical \textit{NewOrder} and \textit{Payment} transactions require modifications by RC+ELM and RC+Promotion, increasing write conflicts on \textit{NewOrder}, resulting in a performance disadvantage compared to \sysname, which can achieve up to 2.3$\times$ their performance. Due to high contention on the warehouse relation, validation overheads are generally higher, except when the proportion of \textit{NewOrder} is 0.1, where \sysname shows a 10.7\% improvement over SI. In other scenarios, \sysname adapts to SI. \textit{Payment} transactions are more write-intensive, prompting \sysname to set the database to RC, which achieves up to 40.6\% performance improvement over SI. Compared to other solutions, \sysname achieves up to 53.5\% performance improvement.

\begin{figure}[]
    \centering
    \begin{minipage}{0.95\linewidth}
        \centering
        \includegraphics[width=\linewidth]{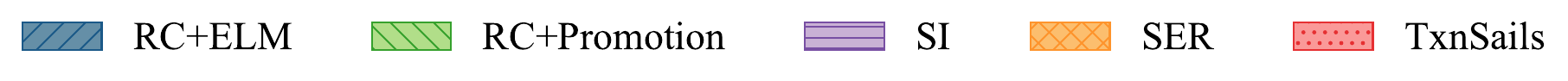}
        \vspace{-5mm}
    \end{minipage}
    \begin{minipage}{0.95\linewidth}
        \centering
        \begin{subfigure}{0.95\linewidth}
            \includegraphics[width=\linewidth]{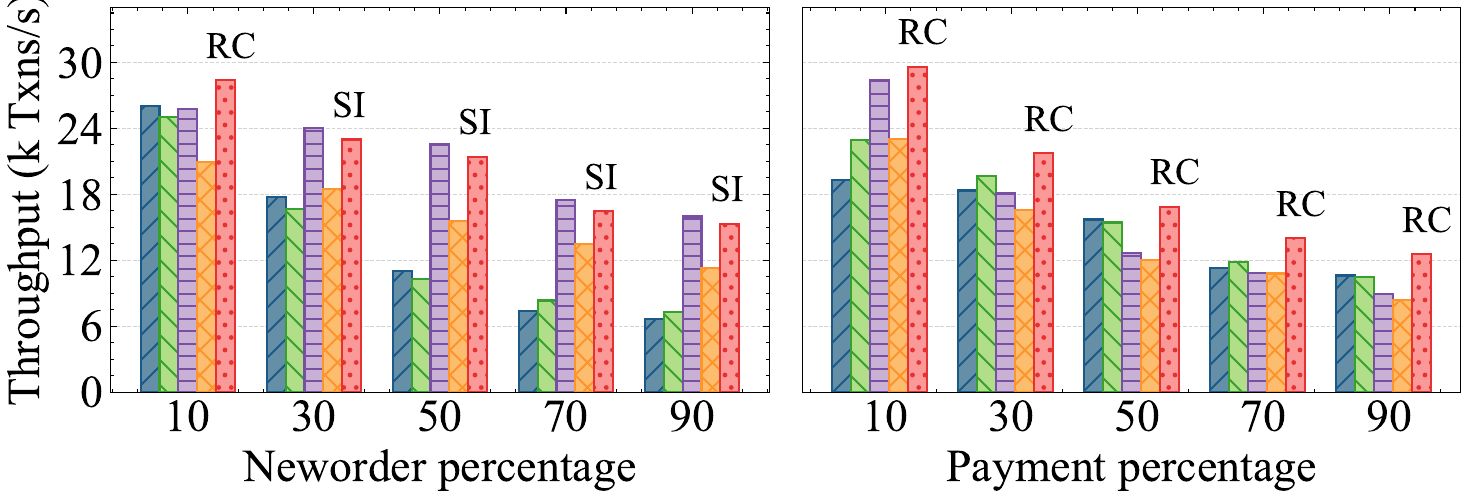}
        \end{subfigure}
    \end{minipage}
    \vspace{-4mm}
    \caption{Impact of templates percentage by TPC-C}
    \label{fig:evaluation.tpcc.ratio}
    \vspace{-4mm}
\end{figure}

\begin{figure}[t]
    \centering
    \begin{minipage}{0.8\linewidth}
        \centering
        \includegraphics[width=\linewidth]{figures/evaluation/bar_legend01.pdf}
        \vspace{-5mm}
    \end{minipage}
    \begin{minipage}{0.95\linewidth}
        \centering
        \begin{subfigure}{0.47\linewidth}
            \includegraphics[width=\linewidth]{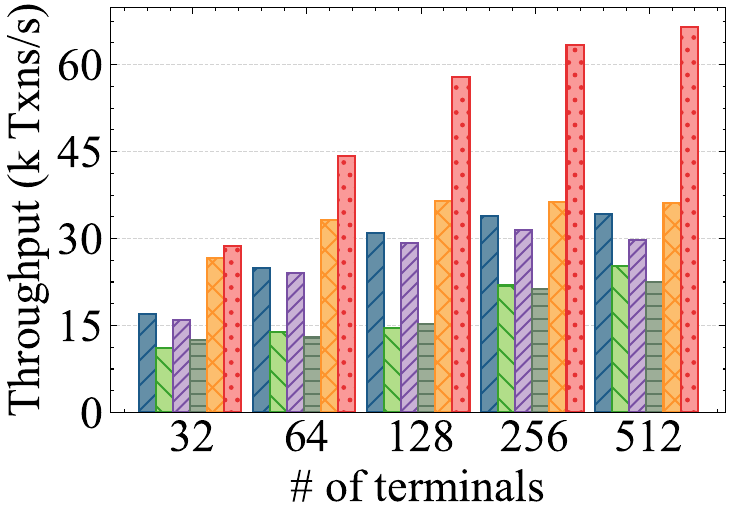}
            \vspace{-6mm}
            \caption{Performance - YCSB}
            \label{fig:ycsb.sca.per}
        \end{subfigure}
        \begin{subfigure}{0.47\linewidth}
            \includegraphics[width=\linewidth]{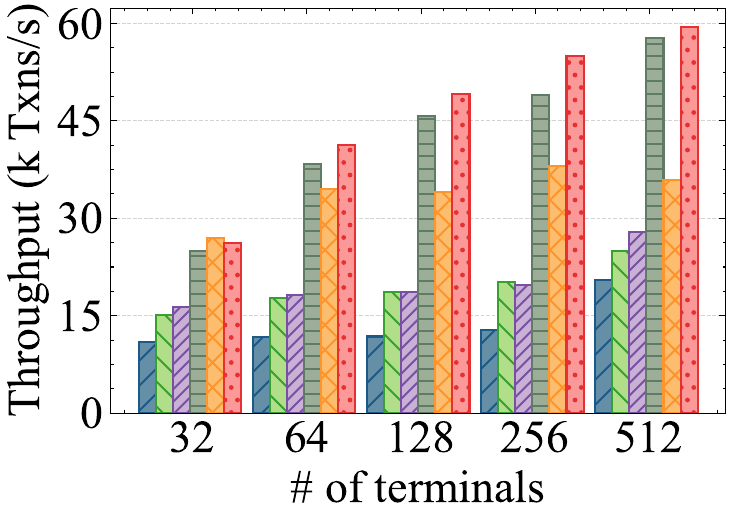}
            \vspace{-6mm}
            \caption{Performance - SmallBank}
            \label{fig:sb.sca.per}
        \end{subfigure}
    \end{minipage}
    \vspace{-4mm}
    \caption{Impact of client terminal numbers}
    \label{fig:evaluation.scalability}
    \vspace{-6mm}
\end{figure}

\subsubsection{Scalability\label{sec:evaluation:scalability}}
This part evaluates the scalability under various numbers of client terminals, as shown in Figure~\ref{fig:evaluation.scalability}.
\sysname outperforms other solutions by up to 3.96$\times$ and 4.21$\times$ by YCSB and SmallBank, respectively. As client terminals increase, \sysname consistently outperforms, primarily due to its small overhead associated with external concurrency control management at a lower isolation level. Interestingly, SER outperforms most other solutions, as these solutions often introduce additional write operations that restrict concurrency and require updating extensive locking information. The notable exception occurs in the SmallBank workload, where the SI+Promotion method surpasses SER. This improvement can be largely attributed to the modification of a limited number of transaction templates within SmallBank.


\noindent \textbf{\underline{Summary.}} 
Current research often limits concurrency and scalability in a coarse-grained manner by replacing read locks with write locks. In contrast, \sysname employs validation-based concurrency control in a fine-grained manner, achieving superior performance compared to state-of-the-art approaches. Furthermore, unlike previous work that merely advocates for a lower isolation level, we argue that, due to the varying structures and proportions of different transaction templates, higher isolation levels can sometimes yield better results, which can be captured and used by \sysname adaptively.

\section{Related Work}
Our study is related to the previous work on concurrency control algorithms that ensure serializable scheduling within and outside the database kernel.

\noindent\textbf{Within the database kernel.} 
Existing works have explored a variety of algorithms to guarantee SER, including 2PL, OCC, timestamp ordering (TO) and their variants \cite{DBLP:journals/pvldb/BarthelsMTAH19, DBLP:journals/csur/BernsteinG81, DBLP:conf/sosp/TuZKLM13,DBLP:conf/sigmod/YuPSD16,DBLP:conf/sigmod/KimWJP16,DBLP:conf/sigmod/LimKA17,DBLP:journals/pvldb/YuXPSRD18, DBLP:journals/tkde/ZhaoZZLLZPD23, DBLP:journals/vldb/WangJFP17, DBLP:journals/vldb/WangJFP18}. 
While these algorithms effectively eliminate anomalies in concurrent transaction execution, they offer varying performance benefits depending on the workload. To address this, some studies propose adaptive concurrency control algorithms for dynamic workloads.
For instance, SMF~\cite{DBLP:journals/pvldb/ChengKCSBCS24:SMF} greedily selects the next transaction based on the one that would result in the least increase in execution time. Tebaldi~\cite{DBLP:conf/sigmod/SuCDAX17:Tebaldi} constructs a hierarchical concurrency control model by analyzing stored procedures.  Polyjuice~\cite{DBLP:conf/osdi/WangDWCW0021:Polyjuice} employs reinforcement learning to design tailored concurrency control mechanisms for each stored procedure, while Snapper~\cite{DBLP:conf/sigmod/Liu00ZS22:Snapper} mixes deterministic and non-deterministic (i.e., 2PL) algorithms.
However, all these algorithms are explicitly tailored for database kernels, which limits their broader applicability and generalizability. In contrast, \sysname requires no kernel modifications and integrates seamlessly with various database systems. More importantly, \sysname boosts performance by adaptively assigning the optimal isolation level based on workload characteristics while preserving SER.

\noindent\textbf{Outside the database kernel.} 
Some application developers prioritize application-level concurrency control using mechanisms such as Java's ReentrantLock or memory stores like Redis \cite{Redis}. Tang et al. \cite{DBLP:journals/tods/WangTZYZGC24, DBLP:conf/sigmod/TangWZYZG022, DBLP:journals/sigmod/TangWZYZG023} provide valuable insights into ad-hoc transactions, which offer flexible and efficient concurrency control on the application side. Bailis et al. \cite{DBLP:journals/pvldb/BailisFFGHS14} introduce the application-dependent correctness criterion known as \textit{I-confluence}, which evaluates whether coordination-free execution preserves application invariants. Conway et al. \cite{DBLP:conf/cloud/ConwayMAHM12} use monotonicity analysis to eliminate the need for coordination in distributed applications.
These techniques demand that developers have a high level of expertise in concurrency control, which can increase the risk of errors. In contrast, \sysname relieves programmers from the complexities of concurrency control, ensuring efficiency through self-adaptive isolation level selection with minimal application modifications.

The most relevant work to ours demonstrates that scheduling entire workloads under low isolation levels can still achieve SER by adjusting specific query patterns. Fekete et al. provide the necessary and sufficient conditions for SI to achieve serializable scheduling~\cite{DBLP:conf/pods/Fekete05, alomari2008serializable}. 
Ketsman et al. \cite{DBLP:journals/tods/KetsmanKNV22, DBLP:journals/pvldb/VandevoortK0N21} investigate the characteristics of non-serializable scheduling under RC and Read Uncommitted isolation levels. This theoretical framework has been further refined with functional constraints by Vandevoort et al.~\cite{DBLP:conf/icdt/VandevoortK0N22}. 
Based on these insights, \sysname can accurately and efficiently achieve serializable scheduling across various isolation levels. To the best of our knowledge, \sysname is the first work to model the trade-off between the performance benefits and the serializability overhead under low isolation levels, achieving the self-adaptive isolation level selection. 

\section{conclusion}
In this paper, we present \sysname, an efficient middle-tier approach that achieves serializability by strategically selecting between serializable and low isolation levels for dynamic workloads. 
\sysname introduces a unified middle-tier validation method to enforce the commit order consistent with the vulnerable dependency order, ensuring serializability in single-isolation and cross-isolation scenarios. Moreover, \sysname considers the trade-off between the performance benefits of low isolation levels and the serializability overhead. It adopts a graph-learned model to extract the runtime workload characteristics and adaptively predict the optimal isolation levels, achieving further performance improvement. 
The results show that \sysname can self-adaptively select the optimal isolation level and significantly outperform the state-of-the-art solutions and the native concurrency control in PostgreSQL. 



\balance

\bibliographystyle{vldb/ACM-Reference-Format}
\bibliography{sample}

\end{document}